\documentclass[12pt]{article}

\usepackage[utf8]{inputenc} 
\usepackage[margin=2.5cm]{geometry} 
\usepackage{lineno}
\usepackage{setspace}
\usepackage{hyperref}
\usepackage{graphicx}  
\usepackage[font={small}]{caption} 
\usepackage{eurosym}
\usepackage{amsmath}
\usepackage{amsfonts}
\usepackage{amssymb}
\usepackage[table]{xcolor}

\title{\textbf{The impact of incorrect social information \\ on collective wisdom in human groups}}

\author{Bertrand Jayles$^{1,2,3,\ast}$\footnote{Corresponding author -- jayles@mpib-berlin.mpg.de}, 
Ram\'on Escobedo$^3$, St\'ephane Cezera$^4$, Adrien Blanchet$^{4,5}$, \\ 
Tatsuya Kameda$^6$, Cl\'ement Sire$^2$, Guy Theraulaz$^{3,5}$ \\}
\date{\small
	$^1$Center for Adaptive Rationality, Max Planck Institute for Human Development, Berlin, Germany \\ 
	$^2$Laboratoire de Physique Th\'eorique, CNRS, Universit\'e de Toulouse, France \\ 
	$^3$Centre de Recherches sur la Cognition Animale, CNRS, Universit\'e de Toulouse, France \\
    $^4$Toulouse School of Economics, INRA, Universit\'e de Toulouse (Capitole), 31000 Toulouse, France \\
    $^5$Institute for Advanced Study in Toulouse, 31015 Toulouse, France \\
    $^6$Department of Social Psychology, The University of Tokyo, 7-3-1 Hongo, Bunkyo-ku, Tokyo 113-0033, Japan \\
     \normalsize
}

\begin{document} 

\maketitle 
\doublespacing
\begin{abstract}
A major problem that results from the massive use of social media networks is the possible spread of incorrect information. 
However, very few studies have investigated the impact of incorrect information on individual and collective decisions. 
We performed experiments in which participants had to estimate a series of quantities before and after receiving social information. 
Unbeknownst to them, we controlled the degree of inaccuracy of the social information through ``virtual influencers'', who provided some incorrect information. 
We find that a large proportion of individuals only partially follow the social information, thus resisting incorrect information. 
Moreover, we find that incorrect social information can help a group perform better when it overestimates the true value, by partly compensating a human underestimation bias. 
Overall, our results suggest that incorrect information does not necessarily impair the collective wisdom of groups, and can even be used to dampen the negative effects of known cognitive biases.
\end{abstract}

\newpage

The digital revolution has changed the way people access and share information.
In particular, the past few decades have seen an exponential increase of media sources and amount of available information~\cite{castells_rise_2009}.
Moreover, a growing distrust in traditional media has given an increasing share of news consumption to social networks and other pathways to relay information.
This facilitated and more diverse access to information may arguably enhance people's ability to make informed decisions, but at the same time such an information overload dramatically increases the difficulty to verify information, understand an issue, or make efficient decisions~\cite{schick_information_1990,klingberg_the_2009}. In certain cases, it has also disrupted the relationship between citizens and the truth~\cite{viner_how_2016,lewandowsky_misinformation_2017}, leading to polarized communities unable to listen to each other~\cite{Bessi2015}. 
Recently, the effects of large scale diffusion of incorrect information and fake news on the behavior of crowds have gained increasing interest, because of their major social and political impact~\cite{vicario_the_2016}.
The propagation of false information is also reinforced by the use of social bots simulating the behavior of Internet users~\cite{monsted_evidence_2017}.
In particular, there has been recent evidence that fake news can propagate faster and affect people deeper than true information on Twitter, especially when they carry political content~\cite{vosoughi_spread_2018}. In this context, there is a strong need to understand how the diffusion of incorrect information among group members affects individual and collective decisions. 

To address this issue, we use the experimental framework of estimation tasks, which is highly suitable for quantitative studies on social influenceability~\cite{moussaid_social_2013,mavrodiev_quantifying_2013,madirolas_improving_2015,chacoma_opinion_2015,yaniv_receiving_2004,soll_strategies_2009,luo_social_2018}.
We performed experiments in which subjects had to estimate a series of quantities with varying levels of demonstrability, before and after having received social information. 
The demonstrability of a quantity can be interpreted as the amount of prior information a group has about it. 
To put it in simple terms, it represents the ``difficulty'' to determining the actual value of a quantity, a notion which will be made explicit and quantitative hereafter.
Knowing the individuals' estimates before and after receiving the social information, as well as the value of the social information, we can deduce their sensitivity to social influence. Moreover, by introducing \emph{``virtual influencers''} providing either the true value or some incorrect information -- without the subjects being aware of it -- we control the \emph{quality} of the information provided to the subjects, and quantify its resulting impact on individual and collective accuracy.

To compare estimates of different quantities, it is usually necessary to normalize them. A natural and commonly used normalization consists in dividing estimates by the true value of the quantity of interest~\cite{lorenz_how_2011,jayles_how_2017}.
Here, we show that this normalization is insufficient for comparing quantities of a very different nature, and that the dispersion of estimates must be included in the normalization process, which has hitherto largely been neglected~\cite{mavrodiev_quantifying_2013,madirolas_improving_2015,luo_social_2018,lorenz_how_2011,jayles_how_2017,kerckhove_modelling_2016}. 
We provide an adequate normalization procedure, and discuss its implications in terms of distributions of estimates.
Moreover, we demonstrate that providing a moderate amount of incorrect information to individuals can counterbalance a human tendency to underestimate quantities~\cite{krueger_single_1982,krueger_reconciling_1989,izard_calibrating_2008,scheibehenne_psychophysics_2019}, and thereby improve estimation accuracy.
We also find that when social information contradicts the underestimation bias, a strategy which consists in compromising between one’s opinion and the group’s opinion increases the performance of groups, even when the social information is substantially inaccurate.
However, compromising with inaccurate social information that does not contradict the bias may amplify its deleterious effects, and hamper collective performances.

Finally, we use a modified version of an agent-based model developed in~\cite{jayles_how_2017} to better understand the present results, and to analyze the collective response of human groups to information of which levels of inaccuracy go beyond the values tested in our experiments. 
The model quantitatively reproduces the experimental results, and confirms the counter-intuitive observation that incorrect information can improve a group's performance, in particular when the group underestimates the true value and when the information compensates this bias.

\section*{Experimental design}

180 subjects participated in our experiment.  
20 sessions were organized, in each of which 9 subjects were asked to estimate 32 quantities.
Each quantity was estimated twice: subjects first provided their personal/prior estimate $E_{\rm p}$. 
Next, they received as social information the geometric mean $G$ of the $\tau$ previous estimate(s) in the sequence ($\tau = 1$ or 3), and were then asked to provide a second/final estimate $E_{\rm s}$.  The value of $\tau$ was unknown to the subjects and so was the exact nature of the mean provided.
Moreover, this second estimate $E_{\rm s}$ was used to update the social information for the corresponding subject in the next session. Hence, our experiment  produced  $9\times 32\times 20=5760 $  personal and second estimates, adding up to a total of 11520 estimates.

We controlled the quality of the social information provided to the subjects, without them being aware of it.
To that end, we inserted  in the sequence of 20 final estimates given by the subjects -- unbeknownst to them -- $n = 0$, 5, or 15 artificial estimates. 
These additional estimates correspond to a fraction $\rho = \frac{n}{20+n} = 0 \, \%$, $20 \, \%$, or $43 \, \%$ of ``\textit{virtual influencers}''. 
Each sequence thus consisted of $N = 20 + n = 20$, 25, or 35 estimates overall, among which 20 were estimates given by 20 actual participants, one per session.
The influencers' estimates were introduced at random locations in the sequences.
The value $T_{\rm I}$ of the influencers' estimates provided to the participants was controlled through a parameter $\alpha$, which represents a distance to the true value quantifying the (in)correctness of the  influencers' estimates ($\alpha = -2$, $-1$, $-0.5$, $0.5$, 1, $1.5$, 2, 3).
$\alpha$ was normalized so as to make it independent of the questions and its computation will be described in the Results section.
In each session and for each question, a subject was thus assigned a value of $\rho$, $\tau$ and $\alpha$, and his/her second estimate was a single step in a sequence of 20, 25, or 35 estimates (Supplementary Figure~S1 provides a graphic representation of the protocol). Note that the estimates of the \textit{virtual influencers} are also used  to update the social information which is then provided to the next subjects in the sequence. 

The quantities to estimate were grouped into four categories: visual perception (number or length of objects in an image), population of large cities in the world, daily life facts, and extreme (astronomical or biological/geological) events.
As we will see, the separation into these loosely defined categories is reflected in the collected data.
Three additional questions were asked, which cannot be assigned to any of these categories (see the list of questions in Supplementary Information).
All experimental details are given in the Materials and Methods section.

\section*{Results}

\subsection*{Comparing quantities of very different nature}

Because humans perceive numbers roughly as their order of magnitude~\cite{dehaene_neural_2003,dehaene_log_2008}, the logarithm of estimates is the natural quantity to consider in estimation tasks, especially for large quantities, rather than the actual estimates themselves.
Distributions of estimates have indeed often been found highly right-skewed, while the distribution of their common logarithm is generally much more symmetric~\cite{mavrodiev_quantifying_2013,chacoma_opinion_2015,lorenz_how_2011}.
An important issue in estimation tasks is to find a proper way to normalize and aggregate estimates arising from questions with very different quantitative answers. 
Within studies, how can one aggregate estimates of quantities that differ by several orders of magnitude?
Between studies, how can we compare findings coming from different sets of quantities?

\begin{figure} [h]
		\centering
		\includegraphics[width=1\textwidth]{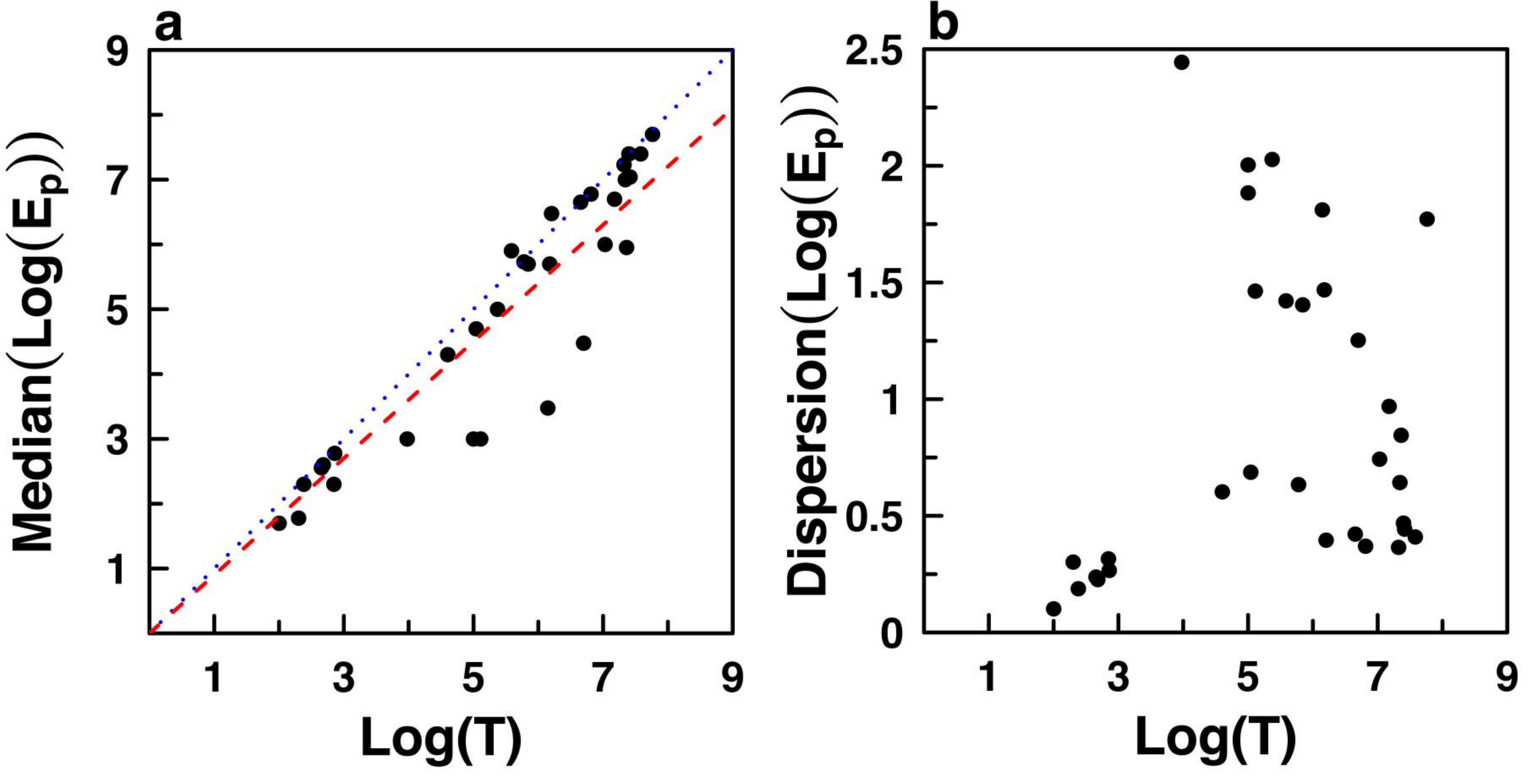}
		\caption{(a) Median and (b) mean dispersion $\langle |\log(E_{\rm p}) - {\rm Median}(\log(E_{\rm p}))| \rangle$ of the logarithms of estimates $E_{\rm p}$, for the 32 questions asked in the experiment. 
		Median($\log(E_{\rm p})$) scales linearly with the log of the true value $T$. The red dashed line is the linear regression, which slope is lower than 1 (blue dotted line), revealing the human tendency to underestimate  quantities.}
		\label{Bias_Disp}
\end{figure}

In line with other works~\cite{kao_counteracting_2018,jayles_debiasing_2020}, we find that the median log-estimate scales linearly with the logarithm $\log(T)$ of the true value (Figure~\ref{Bias_Disp}a), which leads to the natural normalization: $X_{\rm p} = \log \left( \frac{E_{\rm p}}{T} \right)$.
$X_{\rm p}$ represents the deviation of an estimate from the true value in orders of magnitude, and is often used as the quantity of interest in estimation tasks~\cite{mavrodiev_quantifying_2013,madirolas_improving_2015,lorenz_how_2011,jayles_how_2017}.
However, this normalization does not take into account the  dispersion of the log-estimates $\langle |\log(E_{\rm p}) - {\rm Median}(\log(E_{\rm p}))| \rangle$ (where $\langle x\rangle$ refers to the mean of $x$) which can vary considerably for different questions (Figure~\ref{Bias_Disp}b).
In the following, we simply refer to  $X$ as the ``estimates'', dropping the ``log-'' prefix. 

\begin{figure} [h]
	\centering
	\includegraphics[width=1\textwidth]{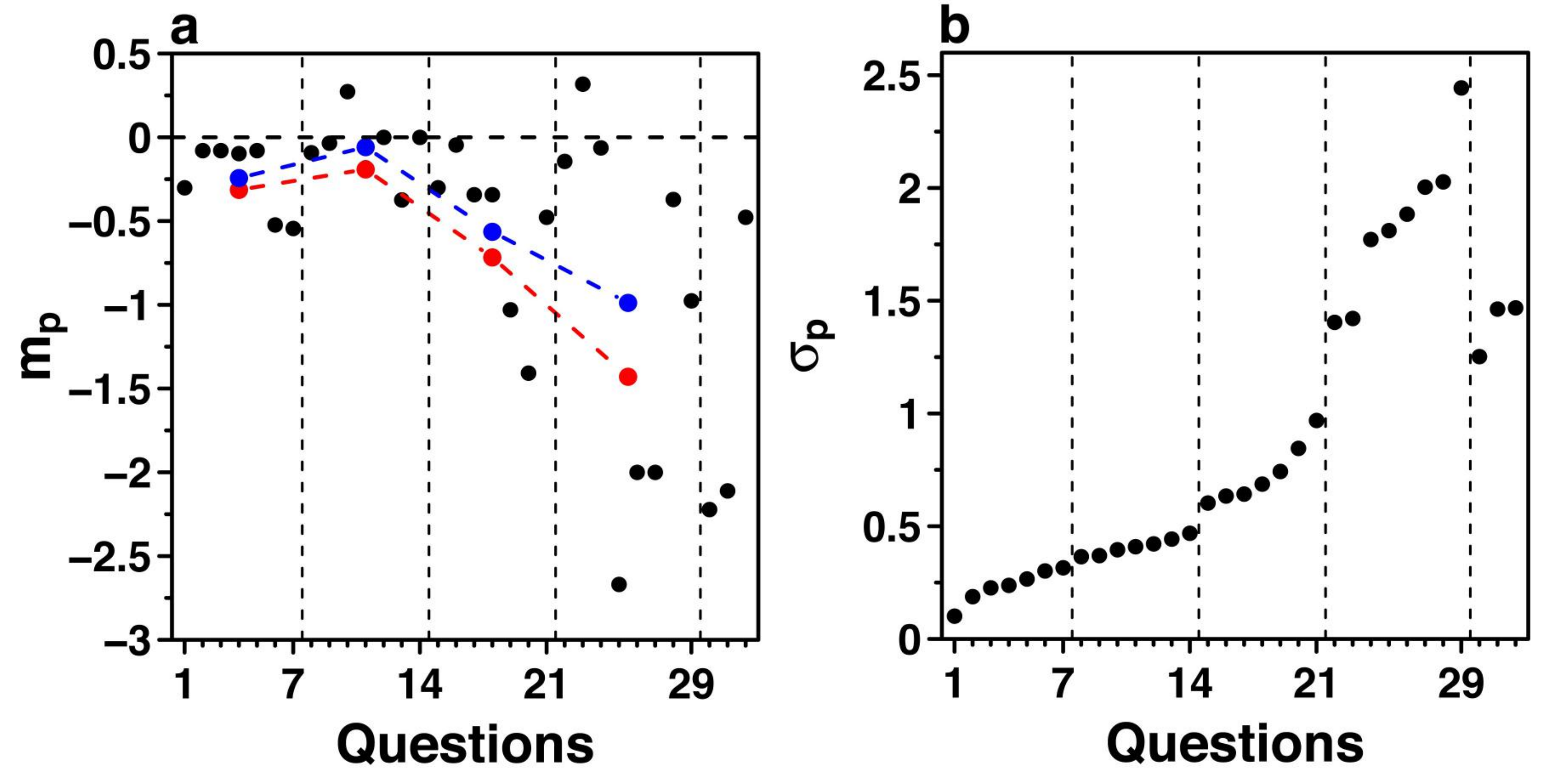}
	\caption{(a) Median $m_{\rm p}$ and (b) dispersion $\sigma_{\rm p} = \langle |X_{\rm p} - m_{\rm p}| \rangle$ of estimates $X_{\rm p} = \log(\frac{E_{\rm p}}{T})$, for the 32 questions asked in the experiment, whose ID are ranked according to their $\sigma_{\rm p}$ which also reflects their demonstrability. 
		The 4 categories of questions (from left to right: visual perception, population of large cities in the world, daily life facts, extreme events), plus the three additional questions, are separated by dashed lines.
		The categories are well distinct in panel (b), indicating that $\sigma_{\rm p}$ is characteristic of the type of quantity to estimate, and more precisely of a question's demonstrability.
		In panel (a), the correlation with demonstrability is much less clear, although $m_{\rm p}$ tends to grow on average when the demonstrability decreases (i.e., when the question ID and $\sigma_{\rm p}$  increase). The blue and red dashed lines in (a) are respectively the average value of $m_{\rm p}$, and the quantity $-\sqrt{\langle m_{\rm p}^2 \rangle}$, in each category.
	}
	\label{m_s_q}
\end{figure}

Figure~\ref{m_s_q}a presents the median $m_{\rm p}$ and Figure~\ref{m_s_q}b the dispersion $\sigma_{\rm p} = \langle |X_{\rm p} - m_{\rm p}| \rangle$ of the personal estimates $X_{\rm p}$, for all questions asked in this experiment (sorted by category of questions). 
One can notice the extreme variation of both quantities depending on the question, suggesting that
including $m_{\rm p}$ and $\sigma_{\rm p}$ in the normalization process is crucial to compare quantities of a different nature. 
Figure~\ref{m_s_q}b shows that the category of a question is clearly identifiable by the dispersion of estimates $\sigma_{\rm p}$ (but not by the median $m_{\rm p}$, see Figure~\ref{m_s_q}a). The natural classification that we have chosen a priori is thus reflected in the experimental data. Moreover, we see that the less demonstrable a question is, the higher the dispersion of estimates.
This is further supported by the three unclassified questions (30 to 32): one could have predicted that they had a low demonstrability (i.e., that people have little prior information about them), and that they would therefore be closer to the ``extreme events'' category than to the other categories, as observed.

\subsection*{Full normalization of estimates}

In a previous study, we found and justified that the estimates $X_{\rm p}$ for low demonstrability questions have a probability distribution function (PDF) close to the Cauchy distribution~\cite{jayles_how_2017}.
This property can be explained by a simple probabilistic argument: if two people provide estimates $X_1$ and $X_2$ of a quantity about which they have \textit{no information at all}, then the average $\frac{X_1 + X_2}{2}$ of both estimates cannot be a statistically better estimation of the correct answer  $T$.
Hence, this average has necessarily the same probability distribution as $X_1$ and $X_2$, and the only distribution that satisfies such a property is the Cauchy distribution (see also Material and Methods).
Our model based on Cauchy distributions convincingly reproduced the experimental data, and in particular fitted well with the experimental distribution of estimates $X_{\rm p}$~\cite{jayles_how_2017}.

\begin{figure} [!h]
		\centering
		\includegraphics[width=0.5\textwidth]{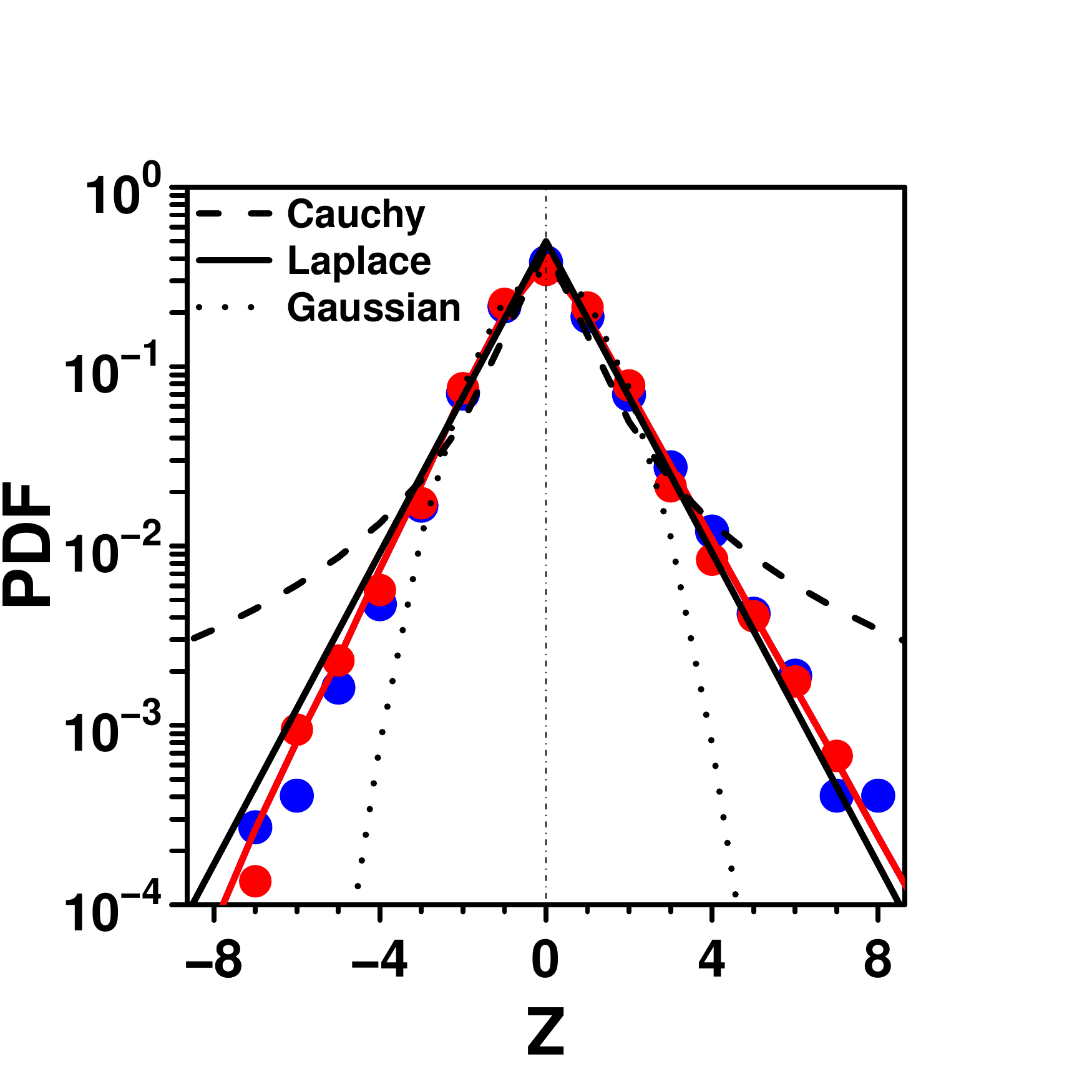}
		\caption{
		Distribution of fully normalized estimates $Z = \frac{X - m}{\sigma}$, before (blue) and after (red) social influence. $m$ and $\sigma$ are respectively the median of the estimates $X = \log(\frac{E}{T})$ and their dispersion, for each corresponding question. $E$ are the actual estimates and $T$ the true value for each corresponding quantity.
		The black lines are the standard (center 0 and width 1) Laplace distribution (full line), the Cauchy distribution (dashed line) and the Gaussian distribution (dotted line) of same width. 
		The Laplace distribution fits the experimental data the best.
		Red lines (overlapping blue lines) are model simulations. 
		}
		\label{Dist_Z}
\end{figure}

However, as we pointed out above, $m_{\rm p}$ and $\sigma_{\rm p}$ have to be considered to compare estimates for questions with answers spanning several orders of magnitude. 
Hence, for each question characterized by its intrinsic median  $m_{\rm p}$ and width $\sigma_{\rm p}$, we normalize the estimate as $Z_{\rm p} = \frac{X_{\rm p} - m_{\rm p}}{\sigma_{\rm p}}$.
Figure~\ref{Dist_Z} shows that the normalized estimates $Z_{\rm p}$ follow the standard Laplace distribution (i.e., with center 0 and width 1), $f(Z)=\exp(-|Z|)/2$, implying that the $X_{\rm p}$ are also Laplace distributed for \textit{individual questions}. 
It is only when different questions with arbitrary widths $\sigma_p$ are aggregated without our normalization that an overall Cauchy-like distribution for the $X_{\rm p}$ emerges.
Similarly, note that after social influence (red dots), the $Z_{\rm s} = \frac{X_{\rm s} - m_{\rm s}}{\sigma_{\rm s}}$, with $m_{\rm s} = {\rm Median}(X_{\rm s})$ and $\sigma_{\rm s} = \langle |X_{\rm s} - m_{\rm s}| \rangle$ also follow the standard Laplace distribution, implying that the $X_{\rm s}$ also follow a Laplace distribution for each question.
We will therefore slightly modify the model developed previously~\cite{jayles_how_2017}, to replace Cauchy distributions by Laplace distributions (see Model section).

By measuring $m_{\rm p}$ and $\sigma_{\rm p}$ and using them in the normalization process, we fix the quantity $\langle |Z_{\rm p}| \rangle = 1$, and therefore have \textit{some} information about the distribution, instead of \textit{none} for the Cauchy distributions argument presented above. As shown in the Material and Methods by exploiting the principle of maximum entropy, the most likely distribution satisfying such a constraint is the Laplace distribution.

This constraint on the dispersion of estimates can be understood as an intrinsic property of the system \{group of individuals, question\}: the dispersion is characteristic of a given group of individuals estimating a given quantity, and gives the typical range of answers that would seem reasonable to most people in the group for that question.
The lower the demonstrability of a question (i.e., the lower the amount of prior information held by individuals in a group about that question), the larger this range.
This is intuitive when considering the following example: an estimate three orders of magnitude from the true value would seem absurd if one considers the age of death of a celebrity, while it would seem perfectly plausible if one considers the number of stars in the universe. 
While the normalization by $m_{\rm p}$ is somewhat trivial (it simply shifts the center of the distribution of $X$ to 0 for every question), the normalization by $\sigma_{\rm p}$ is crucial in order to be able to properly compare and aggregate estimates from different questions (and possibly, from different studies). 
We wish to insist on the fact that this prescription is not a mere methodological detail and that it should be adopted by future works in the field.

In Figure~S2, we show the distribution of $Z$ for the four categories of questions.
One can notice that for very large quantities (Figures~S2c and~d), the left side of the distribution collapses faster than the right side, suggesting that people have an intuition that such quantities must be large, even though they know little about them, such that very small estimates are less frequent.
Such asymmetric Laplace distributions can also be derived from the principle of maximum entropy, by adding a constraint that penalizes small or large estimates (see Materials and Methods).


\subsection*{Model}

\begin{figure} [!h]
		\centering
		\includegraphics[width=1\textwidth]{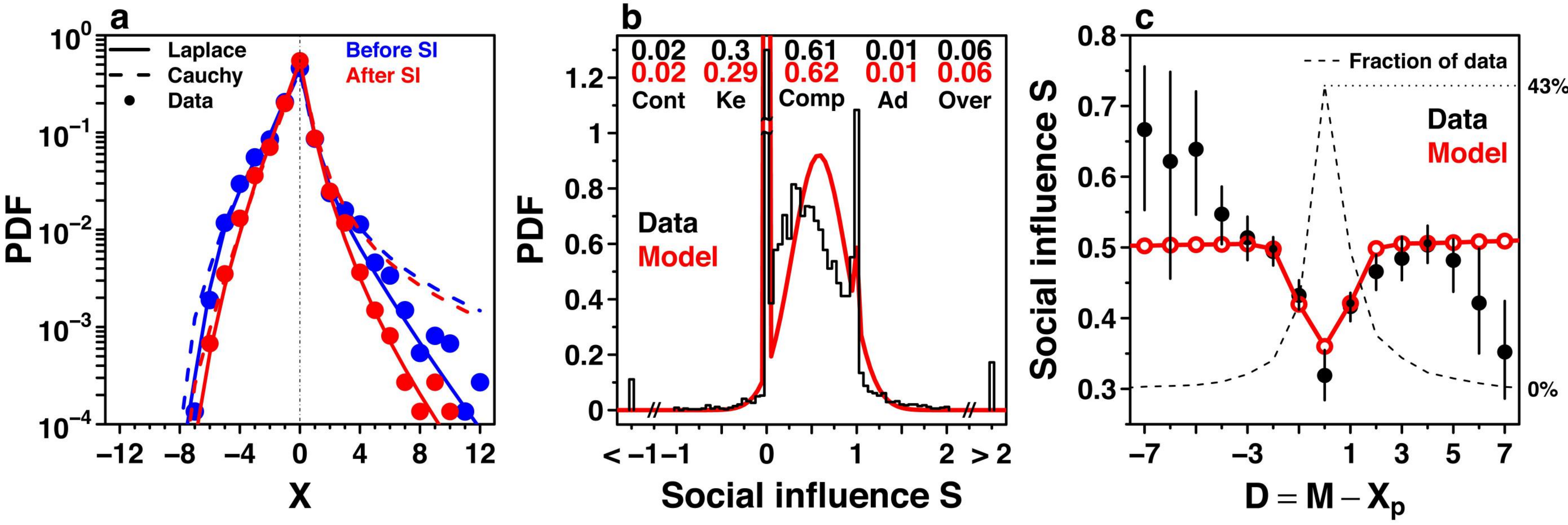}
		\caption{(a) Probability density function (PDF) of individual estimates, before ($X_{\rm p}$, blue) and after social influence ($X_{\rm s}$, red). Dots show experimental data, dashed lines are model simulations based on Cauchy distributions and full lines model simulations based on Laplace distributions.
		Note the sharp decay on the left side of the distribution, well reproduced by the model. The questions asked in our experiment imposed answers higher than one, which translates into $X > - \log(T)$. 
		(b) PDF of the sensitivity to social influence $S$. The fractions of the five behavioral categories are shown, from left to right: contradicters (``Cont'', $S < 0$), keepers (``Ke'', $S = 0$), compromisers (``Comp'', $0 < S < 1$), adopters (``Ad'', $S = 1$) and overreacters (``Over'', $S > 1$).
		Experimental data are shown in black, and model simulations in red. 
		(c) Average sensitivity to social influence $S$ against the distance $D = M - X_{\rm p}$ between the social information $M$ and the personal estimate $X_{\rm p}$. Because the average is sensitive to extreme values, we excluded the values such that $|S| > 100$, which represent less than $1\%$ of the data.
		Black dots correspond to the experimental data, and red empty circles to the model simulations. 
		The dashed line shows the fraction of data for each dot.
		Beyond three orders of magnitude ($|D| > 3$) are only about $7.6 \, \%$ of the data, such that we neglect, in the model, the asymmetric weighting of social information observed in this range of values.
		}
		\label{fig_model}
\end{figure}

In~\cite{jayles_how_2017}, we have introduced an agent-based model to better understand the effects of individual sensitivity to social influence, and of the quantity of information delivered to the individuals, on collective performance and accuracy observed at the group level in estimation tasks.
Estimates $X_{\rm p}$ are drawn from Laplace distributions, the center and width of which are respectively the median $m_{\rm p}$ and dispersion $\sigma_{\rm p} = \langle |X_{\rm p} - m_{\rm p}| \rangle$ of the experimental personal estimates $X_{\rm p}$ for each question.
Figure~\ref{fig_model}a presents the distribution of estimates $X$ for all questions combined, before (blue) and after social influence (red), as well as the corresponding distributions generated by our model, when the $X_{\rm p}$ are generated from Cauchy distributions (as in our previous research~\cite{jayles_how_2017}, dashed lines) and Laplace distributions (full lines). 

The Laplace distribution is able to capture the estimates far from the truth ($X_{\rm {p,s}} > 5$) better than the Cauchy distribution.
It is important to mention that in our previous study~\cite{jayles_how_2017}, the range of possible answers were limited to plus or minus 3, 5, or 7 orders of magnitude from the true value, depending on the question.
By not allowing extreme answers, we probably increased artificially the probability of estimates in the interval [5,7], making the distribution even closer to a Cauchy distribution.

After providing its personal (log-)estimate $X_{\rm p}$, each agent receives as social information the arithmetic mean $M$ of the $\tau$ previous final estimates in the sequence, among which some information $V$ (provided by the \textit{virtual influencers}) is introduced with probability $\rho$. Note that the actual participants where provided the geometrical mean $G$ of the $\tau$ previous  estimates. In terms of log-estimates, the social information $M=\log (G)$ indeed transforms into the standard arithmetic mean.
The agent then provides a second estimate $X_{\rm s}$, defined as the weighted average
of its personal estimate $X_{\rm p}$ and the social information $M$: $X_{\rm s} = (1-S) \, X_{\rm p} + S \, M$, where $S$ is the weight given to the social information, that we call \textit{sensitivity to social influence}.
$S$ can thus be expressed as $S = \frac{X_{\rm s} - X_{\rm p}} {M  - X_{\rm p}}$. 
In Figure~\ref{fig_model}b, we show the distribution of  $S$ from which five natural behavioral categories can be identified: subjects keep their opinion (``keepers'', $S = 0$), compromise with the social information (``compromisers'', $0 < S < 1$), adopt the social information (``adopters'', $S = 1$), contradict it (``contradicters'', $S < 0$) or overreact to it (``overreacters'', $S > 1$). 
In the model, after receiving the social information, an agent keeps its personal estimate ($S = 0$) with probability $P_0$, adopts the social information ($S = 1$) with probability $P_1$, or draws an $S$ in a Gaussian distribution of center $m_{\rm g}$ and width $\sigma_{\rm g}$ with probability $P_{\rm g} = 1 - P_0 - P_1$.

Figure~\ref{fig_model}c shows that the average sensitivity to social influence $S$ increases linearly with the distance $D = M - X_{\rm p}$ between the average social information $M$ and the personal estimate $X_{\rm p}$.
This is implemented in the model by making the probability $P_{\rm g}$ increase linearly with $D$, according to the equation: $\langle S \rangle = P_1 + P_{\rm g} \, m_{\rm g} = a + b \, |D|$, where the intercept $a$ and the slope $b$ characterize the linear cusp observed in Figure~\ref{fig_model}c. More details can be found in the Materials and Methods section.
Notice the subjects' tendency to give more weight to social information that is much lower than their personal estimate ($D < -3$), than to social information that is much higher ($D > 3$). Since this concerns only about $7.6 \, \%$ of the data, we neglect this effect in the model.

The model, originally developed in~\cite{jayles_how_2017}, predicted that by providing subjects with information that overestimates the truth, it was possible to improve individual and collective accuracy more than by providing them with the truth itself, by partly compensating the underestimation bias.
We provide empirical evidence for this prediction in the next section.

Note that the distribution of $X$ narrows after social influence (red dots and lines in Figure~\ref{fig_model}a), implying that estimates have overall gotten closer to the truth, all conditions mixed. 
This may seem counter-intuitive, since in most conditions, incorrect information was provided into the sequence of estimates.
To understand this result, we next investigate the impact of incorrect information on estimation accuracy for each condition separately.

\subsection*{Impact of incorrect information on estimation accuracy}

As explained above, we controlled the quality of the social information received by the individuals, by introducing $n=0$, 5, or 15 \textit{virtual influencers} providing artificial estimates of value $T_{\rm I}$ randomly inserted in the sequences of 20 estimates provided by the participants, and hence affecting the social information delivered to them. 
Since we are looking for an information parameter that is independent of the questions, we define, consistently with the previous discussion on the normalization procedure, the normalized (log) deviation from the truth $\alpha = \frac {\log \left( \frac{T_{\rm I}}{T} \right)} {{\sigma_{\rm p}}_{\rm exp}} = \frac {V} {{\sigma_{\rm p}}_{\rm exp}}$ as an indicator of information quality, where ${\sigma_{\rm p}}_{\rm exp}$ is an expected value of the dispersion of personal estimates $X_{\rm p}$ (the values of the ${\sigma_{\rm p}}_{\rm exp}$ are given in Supplementary Table~S1), and $V$ the (log) deviation  from the truth of the \textit{virtual influencers} estimates $T_{\rm I}$.
We obviously did not know the dispersion of estimates before running the experiment. 
Yet, since the questions were similar to others used in a previous study~\cite{jayles_how_2017}, we could formulate reasonable expectations. 
Indeed, Supplementary Figure~S3 shows that ${\sigma_{\rm p}}_{\rm exp}$ scale more or less linearly with the actual dispersion of estimates $\sigma_{\rm p}$, although it tends to underestimate it.

\begin{figure} [h!]
\centering
\includegraphics[width=1\textwidth]{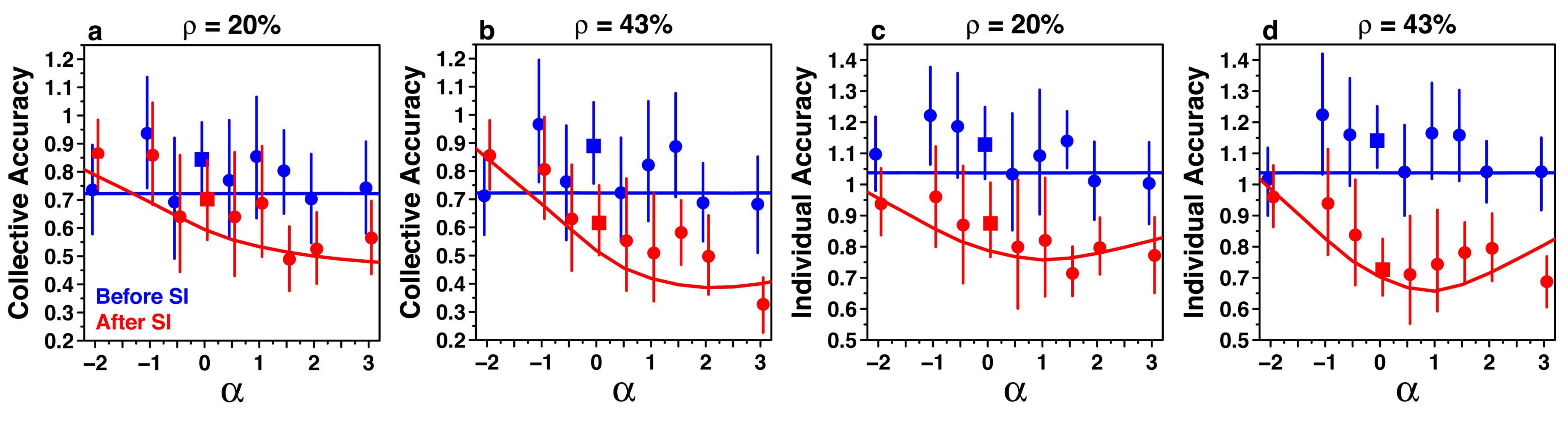}
\caption{Collective (a and b) and individual (c and d) accuracy, as a function of the quantifier of information quality $\alpha$, before (blue) and after (red) social influence, for $\rho = 20 \, \%$ (a and c) and $\rho = 43 \, \%$ (b and d) of influencers in the sequence of estimates. 
Dots are experimental data from the experiment presented here, while squares at $\alpha = 0$ are experimental data from a previous study, in which the same percentage of \textit{virtual influencers} provided some perfectly accurate information~\cite{jayles_how_2017}. 
		Full lines are model simulations.  Surprisingly, incorrect information can be beneficial to collective and individual accuracy, which optima are reached for positive values of $\alpha$, i.e., for incorrect information that overestimates the truth.
The computation of the error bars is explained in the Materials and Methods.}
		\label{PA_VS_alpha}
\end{figure}

$\alpha$ represents the deviation of $T_{\rm I}$ from the truth $T$ in the (expected) natural scale of each question, which is the dispersion ${\sigma_{\rm p}}_{\rm exp}$ of individual errors for that question ($X_{\rm p}$ are deviations from the truth or errors).
The  value $T_{\rm I}$ introduced in the sequence of estimates is hence $T_{\rm I} = T . 10^{\alpha {\sigma_{\rm p}}_{\rm exp}}$, and equals the true value $T$ when $\alpha = 0$. 
Subsequently, to study the impact of information quality on the group performance, we introduce the variable $Y = \frac{X}{\sigma_{\rm p}}$, where $\sigma_{\rm p}$ is the dispersion of  $X_{\rm p}$ for a given question, and define:
\begin{itemize}
	\item \textit{\textbf{individual accuracy}} as the median of the absolute values of the $Y$ of all individuals $i$, averaged over all questions $q$: $\langle {\rm Median}_i(|Y_{i,q}|) \rangle_q$, and
	\item \textit{\textbf{collective accuracy}} as the absolute value of the median of the $Y$ of all individuals $i$, averaged over all questions $q$: $\langle |{\rm Median}_i(Y_{i,q})| \rangle_q$.
\end{itemize}
The individual accuracy measures how close individual estimates are to the truth (i.e., close to 0 in terms of log variables $X$) on average, while the collective accuracy measures how close the median estimate is to the truth.
Figure~\ref{PA_VS_alpha} shows that both measures improve after social influence (i.e., red dots are closer to 0 than blue dots), \textit{over almost the whole range of the considered values of $\alpha$}, suggesting that incorrect information can, counterintuitively, be beneficial to the performance of groups. 
Our results also suggest that individual accuracy slightly improves after social influence when $\rho = 0 \, \%$ (i.e., no \textit{virtual influencers}), but not collective accuracy, confirming previous findings~\cite{jayles_how_2017}.

Moreover, the optimum value $\alpha_{\rm opt}$ of $\alpha$ at which collective or individual accuracy improves the most is strictly positive, confirming the model prediction that such improvement is maximized not by providing perfectly accurate information to individuals, but information that overestimates the true value.
Such incorrect information partly compensates the underestimation bias, thus bringing second estimates closer to the truth.

Collective accuracy before social influence (blue dots and lines) represents the absolute value of the collective bias of the group, averaged over all questions, i.e., the distance between the median estimate and the truth.
If the value of the collective bias is $\alpha_0 \approx -0.72$, one may naively expect that $\alpha_{\rm opt} = - \alpha_0$ in order to compensate the collective bias and thus optimize collective accuracy.
But, since not all subjects follow the social information fully, one should rather expect $\alpha_{\rm opt} > - \alpha_0$, as supported by the data and model.
 
The fraction $\rho$ of \textit{virtual influencers}  has no significant effect on collective accuracy in the data in Figure~\ref{PA_VS_alpha}.
However, the simulations of the model predict that collective accuracy degrades after social influence either when $\alpha < \alpha_{\rm min} \approx -1.2$ (for both $\rho = 20 \, \%$ and $\rho = 43 \, \%$) or when $\alpha > \alpha_{\rm max} \approx 13.4$ for $\rho = 20 \, \%$ and $\alpha_{\rm max} \approx 7.2$ for $\rho = 43 \, \%$, which corresponds respectively to $\langle {\sigma_{{\rm p}_{\rm exp}}}_q \rangle_q \times \alpha_{\rm max} \approx 0.46 \times 13.4 \approx 6.2$ and $\langle {\sigma_{{\rm p}_{\rm exp}}}_q \rangle_q \times \alpha_{\rm max} \approx 0.46 \times 7.2 \approx 3.3$ orders of magnitude beyond the true value (see Figure~S4).
The impact of $\alpha$ is therefore not symmetric with respect to its optimum $\alpha_{\rm opt}$: incorrect information that largely overestimates the truth can still be beneficial to collective accuracy, while incorrect information that only moderately underestimate the true value is enough to damage collective accuracy.
The same analysis remains true for individual accuracy, only with different values of $\alpha_{\rm opt}$, $\alpha_{\rm min}$, $\alpha_{\rm max}$, and $\alpha_0$.

\subsection*{Incorrect information and sensitivity to social influence}

\begin{figure} [h!]
	\centering
	\includegraphics[width=\textwidth]{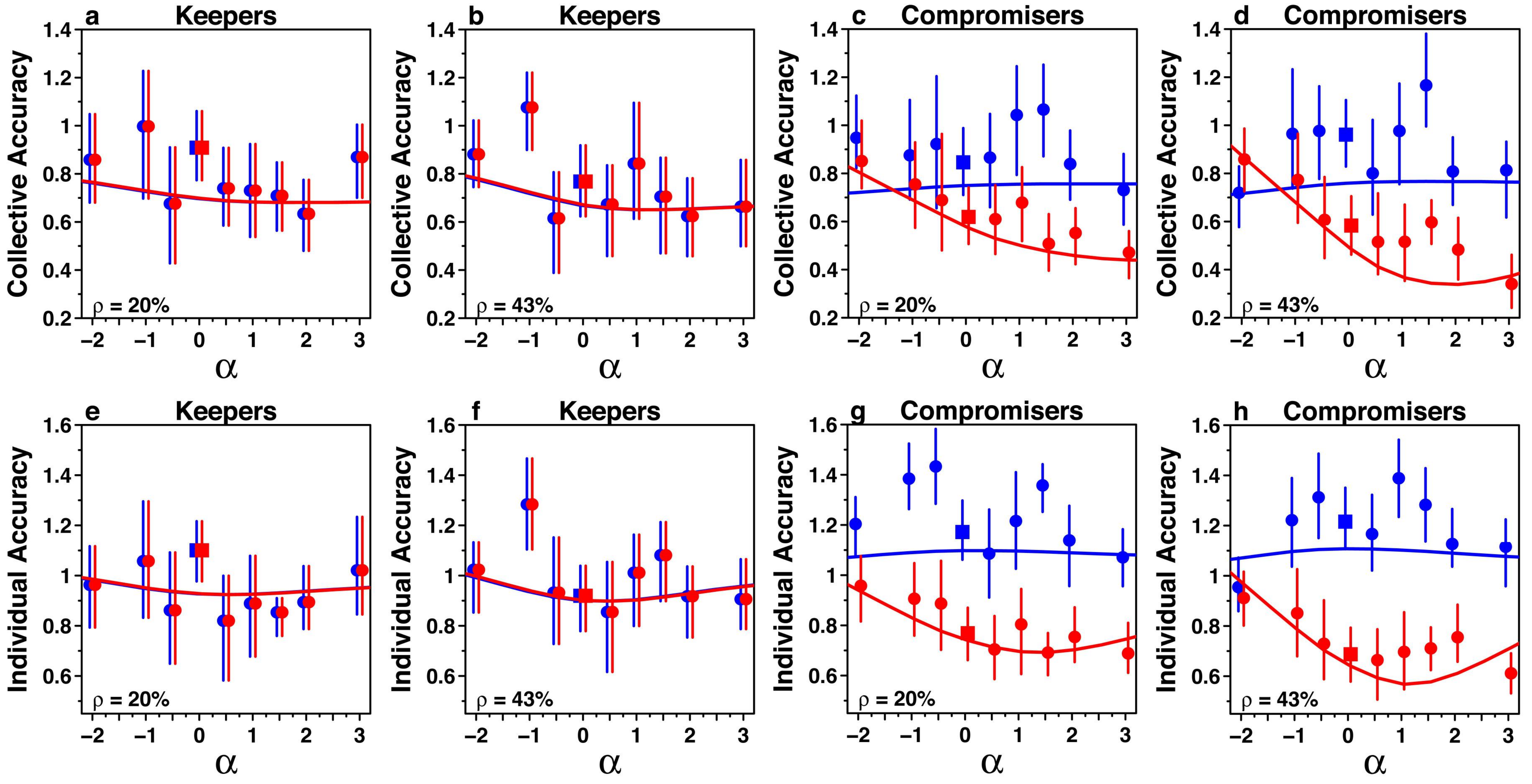}
		\caption{Collective (a, b, c, d) and individual accuracy (e, f, g, h) as a function of the quantifier of information quality $\alpha$, before (blue) and after (red) social influence, for $\rho = 20 \, \%$ (a, c, e, g) and $\rho = 43 \, \%$ (b, d, f, h) of influencers in the sequence of estimates. 
		Keepers ($S=0$) are shown in (a, b, e, f) and compromisers ($0<S<1$) in (c, d, g, h). 
		Dots are experimental data from the experiment presented here, while squares at $\alpha = 0$ are experimental data from a previous study, in which the same percentage of influencers provided some perfectly accurate information~\cite{jayles_how_2017}. 
		Full lines are model simulations. 
		By disregarding social information, keepers are unable to improve in individual and collective accuracy after social influence. 
		Compromisers however, by partly following social information, improve in individual and collective accuracy after social influence, even when influencers provide some incorrect information.}
		\label{IA_VS_alpha_cat}
\end{figure}

It has been shown that estimation accuracy strongly depends on the sensitivity to social influence of individuals in groups~\cite{jayles_how_2017}.
Analyzing the above results in the light of the five behavioral categories of sensitivity to social influence (Figure~\ref{fig_model}b) helps us to understand the mechanisms underlying them.
They cannot be explained by contradicters ($S < 0$), adopters ($S = 1$) or overreacters ($S > 1$), who only represent a small percentage of the population. 
Figure~\ref{IA_VS_alpha_cat} shows collective and individual accuracy as a function of $\alpha$, for the keepers (Figure~\ref{IA_VS_alpha_cat}a) and compromisers (Figure~\ref{IA_VS_alpha_cat}b), which each represents a substantial fraction of the population ($\sim 91 \%$).
Note that the effects are clearer when this separation into behavioral categories is made.

Since keepers disregard social information, we observe no improvement in individual or collective accuracy after social influence (Figure~\ref{IA_VS_alpha_cat}a, b, e, and f).
However, compromisers (Figure~\ref{IA_VS_alpha_cat}c, d, g, and h), who partly follow the social information, significantly improve their performance over the whole range of incorrect information tested here (except for $\alpha = -2$ and $\rho = 43\%$ of \textit{virtual influencers}).
Indeed, because subjects in general, and compromisers in particular, tend to substantially underestimate quantities, they can improve their estimates by following incorrect social information that is closer to the true value than their own personal estimate. 
Moreover, partially following the social information that overestimates the truth allows their second estimates to reach more accurate values, even when the overestimation is quite pronounced. 
Contrariwise, individual accuracy degrades quickly when subjects are given incorrect social information which reinforces their natural cognitive bias by underestimating the true value. 
Compromising thus allows group members to resist incorrect information, as long as this information goes against their cognitive bias.
Similar conclusions can be drawn for collective accuracy, but the patterns are slightly less pronounced.

\begin{figure} [h!]
	\centering
	\includegraphics[width=1\textwidth]{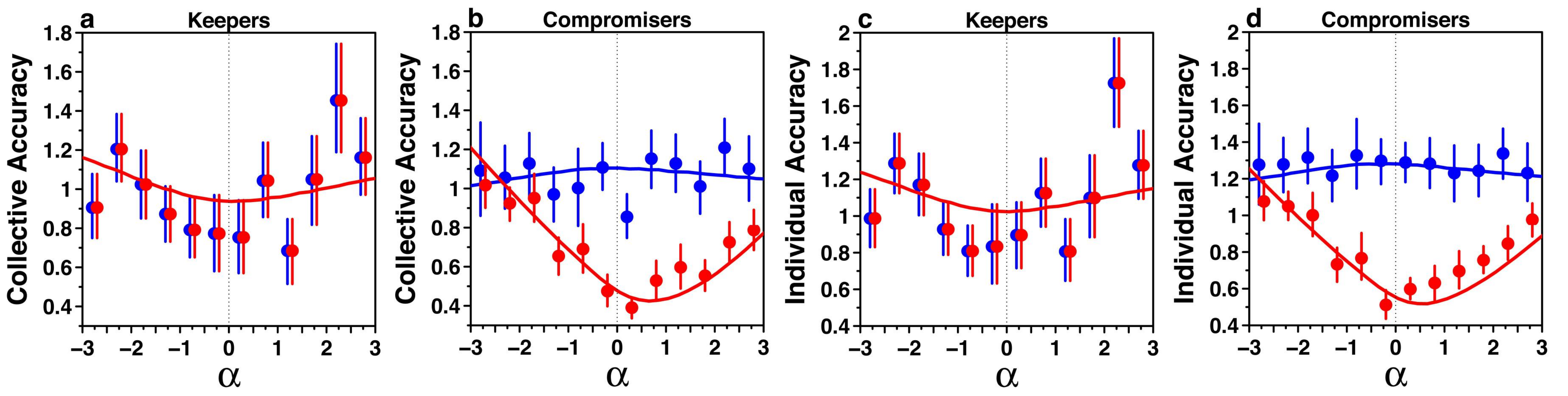}
		\caption{Collective (a, b) and individual (c, d) accuracy against the quantifier of information quality $\alpha$, before (blue) and after (red) social influence, in the separated experiment with the isolated subjects. 
		(a and c) keepers ($S=0$); (b and d) Compromisers ($0<S<1$). 
		Dots are experimental data, and full lines are model simulations.}
		\label{Accu_VS_alpha_extras}
\end{figure}

Figure~\ref{Accu_VS_alpha_extras} shows the equivalent graphs for the ``isolated'' subjects of our experiment (see Materials and Methods).
Isolated subjects received as social information for each question, an estimate $T_{\rm I}$ generated from a \textit{random} value of $\alpha$ uniformly distributed in the interval $[-3,3]$. 

Figure~\ref{Accu_VS_alpha_extras} confirms the above conclusions, but displays sharper patterns, due to a discretization effect:
social information in the main experiment was generated from a discrete set of values of $\alpha$, whereas for isolated subjects, it was drawn from a continuous distribution.

Before social influence (blue), we find that keepers are slightly more accurate than compromisers (average collective accuracy: $0.98$ versus $1.07$; average individual accuracy: $1.08$ versus $1.28$). 
This was already observed in~\cite{jayles_how_2017}, and justified by the fact that a higher tendency to disregard social information is usually associated with a higher average confidence of the subjects in their own estimates, which often comes with a higher prior knowledge about the quantity to estimate.

Note the slight U-shaped curve for keepers in Figure~\ref{Accu_VS_alpha_extras}a and c. This effect is a direct consequence of people's tendency to stick to their personal estimate more when the social information is closer to it (Figure~\ref{fig_model}c): 
when participants receive inaccurate information and retain their opinion, it is often because they were close to it and therefore relatively inaccurate too. Conversely, when participants receive accurate information and keep their opinion, it is often because they were close to it and therefore quite accurate too.
Both effects can be observed in Figure~\ref{IA_VS_alpha_cat}, but are less pronounced there.

\subsection*{Influence of the fraction of virtual influencers on individual behavior}

\begin{figure} [h!]
	\centering
	\includegraphics[width=1\textwidth]{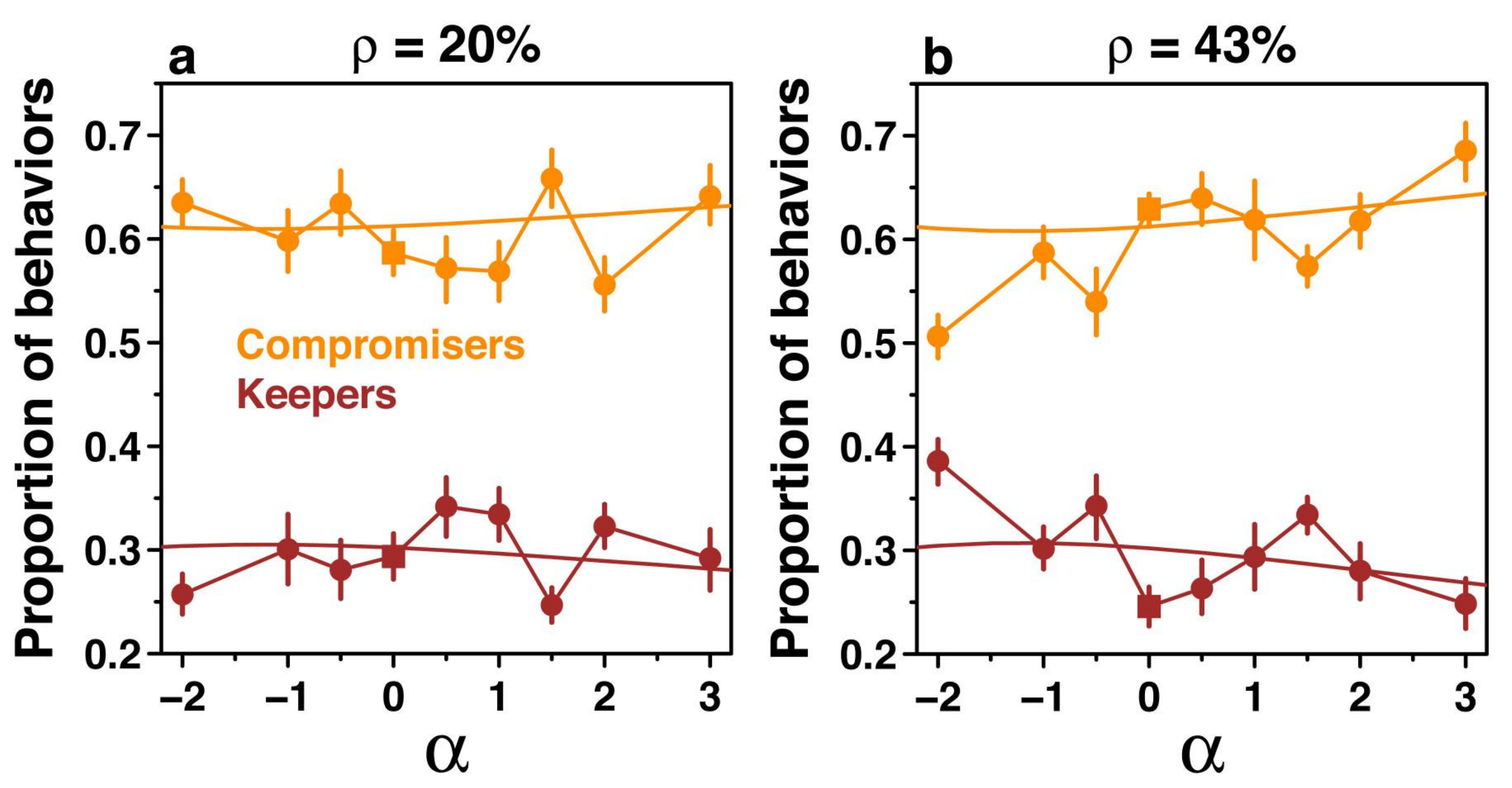}
	\caption{Proportion of compromisers (orange) and keepers (brown) as a function of the quantifier of social information quality $\alpha$, for $\rho = 20 \, \%$ (a) and $\rho = 43 \, \%$ (b) of influencers in the sequence of estimates. 
	Dots are experimental data from the experiment presented here, while squares at $\alpha = 0$ are experimental data from a previous study, in which the same percentage of \textit{virtual influencers} provided some perfectly accurate information~\cite{jayles_how_2017}.
	When $\rho = 43\%$, the fraction of compromisers increases with $\alpha$ at the expense of the fraction of keepers, which decreases with $\alpha$.}
	\label{Prop_VS_alpha}
\end{figure}

We have seen that compromisers, by partially following social information,  were able to improve their accuracy over a wide range of incorrect social information.
Figure~\ref{Prop_VS_alpha} shows the fraction of keepers and compromisers as a function of $\alpha$, when $\rho = 20\%$ (Figure~\ref{Prop_VS_alpha}a) and $\rho = 43\%$ (Figure~\ref{Prop_VS_alpha}b) of \textit{virtual influencers} are introduced in the sequence of estimates.

When $\rho = 20\%$, both fractions of compromisers and keepers remain more or less independent of $\alpha$ (Figure~\ref{Prop_VS_alpha}a).
However, when the proportion of \textit{virtual influencers} providing incorrect information is doubled ($\rho = 43\%$, Figure~\ref{Prop_VS_alpha}b), the fraction of compromisers gradually increases (from 0.5 to 0.68, orange line) with $\alpha$ (from $-2$ to 3), at the expense of the fraction of keepers which decreases (from 0.38 to 0.25, brown line).
For this to happen, this increasing transition from keeping to compromising behavior as $\alpha$ increases thus necessitates a significant proportion of subjects to be provided with incorrect social information. Moreover, this result also suggests that subjects not only adapt their behavior to the degree of incorrectness of the social information they receive, but also tend to compromise more with some social information that overestimates the truth, than with some social information that underestimates it.
The model predicts this increased behavioral transfer with $\alpha$, even when $\rho = 20\%$. This is a direct consequence of the cusp relationship between the sensitivity to social influence $S$ and the distance to the social information $D$ (Figure~\ref{fig_model}c): people tend to compromise with the social information more as it gets farther from their personal estimates (and from the truth) on average.
However, this effect is significantly stronger in the data, suggesting that other mechanisms exist, that are not implemented in the model.

Supplementary Figure~S5 demonstrates that this increasing fraction of compromisers with $\alpha$ when $\rho = 43\%$ leads to an increased improvement in individual and collective accuracy after social influence (but not when $\rho = 20\%$).

\section*{Discussion}

Understanding the effects of incorrect information on individual and collective decisions is crucial in modern digital societies, where social networks and other vectors of information transmission allow a fast and deep flow of information, the accuracy of which is increasingly hard to verify~\cite{centola_how_2018}.
Here, we rigorously controlled the quality of the information delivered to subjects in estimation tasks, by means of ``\textit{virtual influencers}'', i.e., virtual agents inserted into the sequence of estimations -- unbeknownst to the subjects -- and providing a value whose level of inaccuracy was controlled.  
We were thus able to precisely quantify the impact of information quality on individual and collective accuracy in those tasks.

We demonstrated that a proper normalization of estimates must take into account their dispersion, which gives the natural range of ``reasonable'' estimates of a given quantity for a given group.
This normalization process led to the conclusion that estimates follow a Laplace distribution when subjects have little prior information about a quantity to estimate.
Early research showed that in many data sets, estimates $X$ (i.e., deviations from the truth) were often close either to Gaussian distributed or to Laplace distributed~\cite{Laplace_1774,wilson_first_1923}.
Later work have encompassed Laplace and Gaussian distributions into a broader family of exponential distributions, the Generalized Normal Distribution (GND) family~\cite{rider_1924,nadarajah_a_2005}, of PDF:

\begin{align}
	f(X,m,\sigma,n) = \frac{1}{2 \, \sigma \, \Gamma(1+1/n)} \, {\rm exp} \left\{ - \left| \frac{X-m}{\sigma} \right|^n \right\},
\end{align}
where $m$ is the center of the distribution (often called \textit{location} parameter), $\sigma$ is the width of the distribution (often called \textit{scale} parameter) and $n$ is the tailedness (often called \textit{shape} parameter), which controls the thickness of the tails. 
The fatter the tails of a distribution, the higher the probability to find outliers (i.e., estimates that are very far from the distribution center). 
More recent work has studied various data sets of estimates and forecasts in the light of Generalized Normal Distributions, and showed that the tailedness of distributions ranged from $n=1$ (Laplace distribution) to $n=1.6$, $n$ being equal to 2 for Gaussian distributions~\cite{lobo_human_2010}. 
They concluded that most distributions of estimates for usual quantities are actually closer to Laplace distributions than to Gaussian distributions. 
This discussion can be related to the amount of prior information held by a group about a certain quantity. 
We found that when a quantity is ``hard'' to estimate (i.e., low demonstrability, corresponding to a low amount of prior information about the quantity in the group), the expected distribution of estimates is very close to a Laplace distribution. 
When a quantity is ``easy'' to estimate (i.e., high demonstrability, corresponding to a high amount of prior information about the quantity in the group), few  outliers are expected, such that the distribution of estimates could be expected to be closer to a Gaussian distribution. 
However, our results show that regardless of the questions demonstrability, distributions of estimates are significantly closer to Laplace distributions than to Gaussian distributions when properly normalized, in agreement with~\cite{lobo_human_2010}.
In any case, we consider that future studies involving estimation tasks should apply the normalization procedure presented here when  comparing and  aggregating the estimates of different quantities, for which the width $\sigma_{\rm p}$ should be used to quantify their demonstrability.

We then studied the impact of incorrect information on individual and collective accuracy, and found that providing incorrect information that overestimates the true value can help a group perform better than providing the correct value itself, by partly compensating for the human underestimation bias.
Moreover, collective and individual accuracy can improve after social influence over a surprisingly wide  range of incorrect information.
This counter-intuitive result is a consequence of a large proportion of individuals compromising with the social information, i.e., partially following it.
By doing so, subjects are able to benefit not only from relatively accurate social information, but also from incorrect information that goes against their cognitive bias. 
Indeed, because of the human tendency to underestimate quantities, partially following an overestimation of the truth -- even a large one -- can bring second estimates closer to the truth, thus improving accuracy.
However, incorrect social information can also harm accuracy if it amplifies the bias. 
This may be related to some deleterious effects of social information observed at times, for instance, how the spread of misinformation can deeply affect the behavior of crowds as well as public opinion~\cite{karlova_social_2013,monacu_collective_2015,qiu_limited_2017}.

In a former study~\cite{jayles_how_2017}, we showed that adopting the social information was the best strategy in order to improve accuracy, if \textit{virtual influencers} provide perfectly accurate information in the sequence of estimates. 
However, while adopting can lead to higher performance than compromising in this particular case, our results show that compromising offers more resilience when the information provided is potentially less accurate. 

We also found that subjects were sensitive to the degree of incorrectness of the social information they received. 
They adapted their behavior to the social information, by compromising more with the social information as it overestimated the truth more, and compromising less as it underestimated the truth more.
This asymmetric strategy is surprisingly well adapted to counter the human underestimation bias.
Indeed, as explained above, following (even partly) social information that underestimates the truth may increase the bias, while following social information that overestimates the truth may decrease it.
Following less in the former case, and more in the latter is thus bound to increase the performance of groups.
Former studies have already observed this subjects' tendency to rely more on social information that is higher than their personal estimate, than on social information that is lower, and showed that it had valuable consequences for collective performance in estimation tasks~\cite{kao_counteracting_2018,jayles_debiasing_2020}. 
In~\cite{jayles_debiasing_2020}, it is suggested that people can more easily assess the validity of small numbers compared to large numbers, because they have no direct experience with events related to those large numbers~\cite{resnick_dealing_2017}, and as a consequence reject more often small numbers provided by the social information.

We then used a modified version of a model of collective estimation developed in~\cite{jayles_how_2017}. 
The predictions of the model are in good agreement with the experimental data, and confirm that to optimize collective accuracy, social information must do more than compensate the initial collective bias, as most individuals only partly follow social information.
In addition, the impact of the quality of information is not symmetric with respect to its optimum: collective accuracy can be improved by delivering incorrect information which overestimates the true value by up to several orders of magnitude, whereas it decays fast if the information delivered only slightly underestimates it. In other words, social information reinforcing the bias of the group has a strong negative impact on its accuracy.

Overall, we found that incorrect social information does not necessarily impair the collective wisdom of groups, and can even be used to counter some deleterious effects of cognitive biases.
Individuals demonstrated an ability to discriminate the validity of social information, depending on its distance from their personal estimates, and thus to benefit from accurate social information, while at the same time resisting inaccurate social information.
These results suggest that people may be more resilient to malicious information than is often thought, and at the same time that the negative effects of identified biases can be dampened by exchanging relevant social information, thus improving collective decisions.

\vskip 0.5cm
\noindent \textbf{Data:} The data supporting the findings of this study are available at fig\textbf{share}: \\
https://doi.org/10.6084/m9.figshare.11903421 

\vskip 0.5cm
\noindent \textbf{Funding:} This work was supported by Agence Nationale de la Recherche
project 11-IDEX-0002-02–Transversalité–Multi-Disciplinary Study of Emergence
Phenomena, a grant from the CNRS Mission for Interdisciplinarity
(project SmartCrowd, AMI S2C3), and by Program Investissements d’Avenir
under Agence Nationale de la Recherche program 11-IDEX-0002-02, reference
ANR-10-LABX-0037-NEXT. B.J. was supported by a doctoral fellowship
from the CNRS, and R.E. was supported by Marie Curie Core/Program Grant
Funding Grant 655235–SmartMass. T.K. was supported by Japan Society for
the Promotion of Science Grant-in-Aid for Scientific Research JP16H06324
and JP25118004. 

\vskip 0.5cm
\noindent \textbf{Author Contributions:} B.J., C.S., and G.T. designed research; B.J., R.E., S.C., A.B., T.K., C.S., and G.T. performed research; B.J., C.S., T.K., and G.T. analyzed data; and B.J., C.S., and G.T. wrote the article with critical input from all other authors. 

\vskip 0.5cm
\noindent \textbf{Competing Interests:} The authors declare that they have no competing interests.



\section*{Materials and methods}

\subsection*{Ethics statement}

The aims and procedures of the experiments conformed to the Toulouse School of Economics Ethical Rules. All subjects provided written consent for their participation.

\subsection*{Experimental design}

180 subjects -- mostly students from the University of Toulouse -- participated in our experiments.  
20 sessions were organized, in each of which 9 subjects had to answer 32 questions for which they had first to give a prior/personal estimate and then a second/final estimate, the latter after being confronted to social information. 
Each final estimate in each session constituted a step in a sequence of 20 estimates.
Hence, our main experiment (see the experiment with ``isolated subjects'' hereafter) produced a total of  9 subjects $\times$ 32 questions $\times$ 20 sessions $= 5760$  final estimates and the same number of prior estimates.
For each question, subjects first provided their prior/personal estimate $E_{\rm p}$. 
Next, they received as social information the geometric mean $G$ of the $\tau$ previous estimate(s) in the sequence ($\tau = 1$, 3), and were then asked to provide a second/final estimate $E_{\rm s}$ (see illustration in Supplementary Figure~S1). 
Subjects were oblivious to the actual value of $\tau$, only being told that the social information was the average of the final answers of \textit{some} previous participants.
The time allowed to provide an answer was limited to 40\,s per estimate, after which a blinking text urging the subjects to answer quickly would appear on their computer screen.

The specificity of our experiment lies in the design of a system aimed to control the quantity and quality of the social information provided to the subjects, without them being aware of it.
To that end we inserted, at random locations in the sequence of estimates (and unbeknownst to the subjects), $n = 0$, 5, or 15 artificial estimates, corresponding to a fraction $\rho=n/(20+n) = 0 \, \%$, $20 \, \%$, or $43 \, \%$ of \textit{virtual influencers}. 
Each sequence thus consisted of a total  of $N = 20+n= 20$, 25, or 35 estimates and the estimates of the \textit{virtual influencers}  were also used to compute the updated social information for the following participants in each sequence.
We controlled the value $T_{\rm I}$ of the artificial estimates, through a parameter $\alpha$ ($\alpha =$ -2, -1, -0.5, 0.5, 1, 1.5, 2, 3) defined such that $T_{\rm I} = T . 10^{\alpha {\sigma_{\rm p}}_{\rm exp}}$, where $T$ is the true value of a quantity to estimate, and ${\sigma_{\rm p}}_{\rm exp}$ an expected value of the dispersion of estimates $\sigma_{\rm p}$ to be obtained. 
The closer $T_{\rm I}$ to $T$ (i.e., the closer $\alpha$  to 0), the higher the information quality. In particular, when $\alpha = 0$, artificial estimates are equal to the true value ($T_{\rm I} = T$). 
A proper justification of the choice of $\alpha$ as the quantifier of information quality, and a detailed explanation of its physical interpretation, are provided in the main text.
The range of values of $\alpha$ used in the experiment was chosen by exploiting preliminary simulations of the model developed in~\cite{jayles_how_2017}, that we describe in detail below.

Each participant in each session was associated a value of $\rho$ ($\rho = 0\%$ for subject 1, $\rho = 20 \, \%$ for subjects 2 to 5 and $\rho = 43 \, \%$ for subjects 6 to 9). 
For subjects 2 to 9 ($\rho \neq 0\%$; no $\alpha$ for $\rho = 0\%$), one particular value of $\alpha$ was associated to each question. 
We distributed the values of $\alpha$ such that at the end of each session (20 sessions), each combination of a fraction $\rho$ of influencers ($\rho \ne 0$) and information quality $\alpha$ was repeated 16 times, such that the experimental points in Figures~5, 6 and 8 are all constructed from $16 \times 20 = 320$ values.
For $\rho = 0 \, \%$, there were 32 sequences per session (640 values).
Although our previous study did not show evidence for a significant effect of $\tau$ on the dynamics~\cite{jayles_how_2017}, we decided to perform an additional check.
A value of $\tau$ (1 or 3) was thus associated to each question, independent of the values of $\rho$ and $\alpha$. 
Preliminary results confirmed that the effect of $\tau$ can be neglected, and we thus combined the data from both values of $\tau$ in all graphs presented in this paper.
Supplementary Table~S1 summarizes the conditions $(\rho, \alpha, \tau)$ for each subject and each question in a session. 
In all sessions, these conditions were repeated and the order of the questions was randomized.
In the first session (i.e., the first step of all sequences), subjects received as social information (initial condition) the value $T_{\rm I}$ provided by the influencers. 
In the particular case $\rho = 0 \, \%$ (no influencers), the initial social information provided was the true answer to the question $T$.
Anyway, note that our previous study showed that the choice of the initial condition had little impact on the subsequent dynamics~\cite{jayles_how_2017}.
At the end of each session, subjects were rewarded according to their overall performance/accuracy: 20\euro~for the two first ones, 15\euro~for the next four ones and 10\euro~for the three last ones.

As a safeguard against subjects not showing up, 3 additional subjects were recruited in each session. 
Hence, when more than nine subjects were present in a session, the additional subjects (``isolated subjects''; 51 in total) were given a special treatment, unbeknownst to them: they were not part of a sequence or associated to any specific condition ($\rho$, $\alpha$, $\tau$), and received instead, as social information, a value $T_{\rm I}= T . 10^{\alpha {\sigma_{\rm p}}_{\rm exp}}$ generated from a random value of $\alpha$ uniformly distributed in the interval $[-3,3]$. 
These ``isolated subjects''   were paid 10\euro, independently of their performance.

\subsection*{Cauchy and Laplace distributions} 

In the context of estimation tasks, the Cauchy and Laplace distributions emerge naturally.

\subsubsection*{Cauchy distribution}

The Cauchy distribution centered at $m$ (the median) and of width $\sigma$ is given by
\begin{equation}
f(X,m,\sigma) = \frac{1}{\pi} \frac{\sigma}{(X-m)^2 + \sigma^2}.
\end{equation}
The Cauchy distribution has  fat tails and an infinite variance. One of the notable and unique properties of the Cauchy distribution is that, if $X_1,..., X_N$ are independently drawn from the same Cauchy distribution, then the average $\bar X_n=\sum_i^n X_i/N$ (or any weighted average of the $X_i$) has \textit{exactly the  same} Cauchy distribution as the $X_i$, i.e., with the same center and width. In particular, the distribution of the average $\bar X_n$ does not have a smaller width than the distribution of the $X_i$, as one would obtain if the $X_i$ were drawn from a distribution with a finite variance. 
For instance, if the $X_i$ are drawn in the same Gaussian distribution with standard deviation $\sigma$,  the average $\bar X_n$ is also Gaussian distributed, but with a reduced standard deviation $\sigma_n=\sigma/\sqrt{N}$, a basic property leading to the law of large numbers. 

In the context of estimation tasks, if $N$ subjects having absolutely no clue about the actual answer to a question draw their personal estimates  $X_1,..., X_N$ from some random distribution, one cannot expect the average of the estimates $\bar X_n$ to provide more information than the individual estimates, and $\bar X_n$ should then have the same distribution as the $X_i$. The unique probability distribution satisfying this property is the Cauchy distribution.

\subsubsection*{Laplace distribution}

Now, let us assume that the distribution for a random variable $X$ is known to be symmetric and centered in $m$ with a known width given by $\sigma=\int_{-\infty}^\infty |X-m|f(X)\,dX$. What is the most likely distribution satisfying this property? To answer this question, one has to maximize the entropy $S(f)=-\int_{-\infty}^\infty f(X)\log [f(X)]\,dX$, subject to the above constraint for $\sigma$ along with the normalization constraint  $\int_{-\infty}^\infty f(X)\,dX=1$. This optimization problem (maximum entropy principle) with constraints is solved by introducing two Lagrange multipliers $\alpha$ and $\beta$, and maximizing with respect to $f(X)$ the functional $F(f)$ given by
\begin{equation}
F(f)=-\int_{-\infty}^\infty f(X)\log [f(X)]\,dX+\alpha\int_{-\infty}^\infty f(X)\,dX
+\beta \int_{-\infty}^\infty |X-m|f(X)\,dX.\label{free}
\end{equation}
Expressing that the functional derivative of $F(f)$ with respect to $f(X)$ vanishes (maximum of $F(f)$), one obtains
\begin{equation}
 -\log [f(X)]-1+\alpha+\beta |X-m|=0,
\end{equation}
leading to $f(X)=\exp(1-\alpha-\beta |X-m|)$. Finally, $\alpha$ and $\beta$ are computed by expressing the normalization and width constraints leading to the symmetric Laplace distribution
\begin{equation}
f(X,m,\sigma) = \frac{{\rm e}^{-\frac{ |X-m|}{\sigma}}}{2\sigma}. 
\end{equation}
The normalized variable $Z = \frac{X - m}{\sigma}$ has
the standard Laplace distribution (i.e., with center 0 and width 1), $f(Z)=\exp(- |Z |)/2$.
 
Note that if the assumption of a symmetric distribution is relaxed and that one wants for instance to penalize large or small values of $X$ (in the context of human estimations, this is related to the bias against large answers), one can add a term $\gamma\int_{-\infty}^\infty Xf(X)\,dX$ in $F(f)$, adding an independent constraint on the mean of $X$. The same calculation as above leads to an asymmetric Laplace distribution with different exponential tails below and above the maximum of the distribution, reminiscent of what is observed experimentally in Supplementary Figure~S2c and S2d for estimates about questions with very large answers. 

In physics, when the  variable $X$ is the energy of the considered system, the only constraint in addition to the normalization is the knowledge of its average energy $E=\int_{-\infty}^\infty Xf(X)\,dX$. Repeating the same optimization procedure as above ($F(f)$ in Eq.~(\ref{free}) is then proportional to the opposite of the free energy, to be maximized), one finds that the most likely distribution for the energy of the system is the famous Boltzmann distribution 
\begin{equation}
f(X)=\frac{{\rm e}^{-\beta X}}{\cal Z},
\end{equation}
where $\cal Z$ is a normalization constant, and the Lagrange multiplier $\beta$ can be ultimately shown to be proportional to the inverse of the temperature $\cal T$, $\beta=(k_{\rm B} {\cal T})^{-1}$, where $k_{\rm B}$ is the Boltzmann constant.

\subsection*{Model} 

\subsubsection*{Main experiment}

Our model simulates a sequence of $N = 20 + n$ successive estimates performed by 20 agents and $n$ \textit{virtual influencers} for a given question. 
The model directly implements the log deviations from the truth $X = \log \left( \frac{E}{T} \right)$.
A typical run of the model consists of the following steps, for an agent estimating a given quantity in a given condition $(\rho,\alpha,\tau)$:

\begin{enumerate}
	\item An agent's personal estimate $X_{\rm p}$ is drawn from the Laplace distribution $f(X_{\rm p},m_{\rm p},\sigma_{\rm p})$, where $m_{\rm p}$ and $\sigma_{\rm p}$ are respectively the median and average absolute deviation from the median	of the experimental distribution of estimates for each question. We impose that estimates $E_{\rm p}$ are greater than 1, i.e., $X_{\rm p} > - \log(T)$. 
	This condition explains the fast decay on the left side of the distribution in Figure~\ref{fig_model}a;
	
	\item With probability $\rho = \frac{n}{n+20}$, a \textit{virtual influencer} plays and provides the value $V = \alpha \, {\sigma_{\rm p}}_{\rm exp}$ (the values of ${\sigma_{\rm p}}_{\rm exp}$ for each question are given in Supplementary Table S1). 
	With probability $(1-\rho)$, the agent receives as social information the average $M$ of the $\tau$ previous estimates (possibly including estimates from \textit{virtual influencers}). $M_0 = V$ is used as initial condition for the first agent;
	
	\item The agent's sensitivity to social influence $S$ is drawn from a Gaussian distribution of mean $m_{\rm g} = 0.58$ and standard deviation $\sigma_{\rm g} = 0.3$ with probability $P_{\rm g}$, or takes the value $S=0$ or $S=1$ with probability $P_0$ and $P_1 = 1-P_0-P_{\rm g}$, respectively (Figure~\ref{fig_model}b). 
	$P_{\rm g}$ and $P_0$ have a linear cusp relationship with the distance $D = M - X_{\rm p}$ between the social information $M$ and the personal estimate $X_{\rm p}$ (Figure~\ref{fig_model}c), while $P_1 = 0.012$ is kept independent of $D$.
	
	For a given distance $D$, the average sensitivity to social influence is $\langle S \rangle = P_0 \times 0 + P_1 \times 1 + P_{\rm g} \times m_{\rm g} = a + b \, |D|$, where the intercept $a = 0.34$ and the slope $b = 0.09$ are the coefficients of the linear cusp relationship extracted from Figure~\ref{fig_model}c. $P_{\rm g}$ is hence given by $P_{\rm g} = (a + b \, |D| - P_1)/m_{\rm g}$.
	A plateau is added at $|D| > 3$ by setting an upper limit ${P_{\rm g}}_{\rm max} = 0.85$ on $P_{\rm g}$.

	\item The agent's final estimate ${X_{\rm s}}$ is the weighted average of the personal estimate and the social information: $X_{\rm s} = (1 - S) \, X_{\rm p} + S \, M$. 
	The condition $X_{\rm s} > - \log(T)$ is also imposed on the $X_{\rm s}$. 
	
	\item One starts again from step 1 for the next agent. 

\end{enumerate}

Apart from $\alpha$ over which we loop in order to make it pseudo-continuous and to explore values beyond those tested in the experiment, our model closely follows the experimental structure: one subject is associated to $\rho = 0\%$, four subjects to $\rho = 20\%$ ($5/25$) and four subjects to $\rho \approx 43\%$ ($15/25$). Moreover, each question is associated with the corresponding value $\tau$ used experimentally (see Supplementary Table~S1).
One simulation of the model thus mimics an experimental run, and results are calculated for each simulation, then averaged over 10000 simulations.

\subsubsection*{Isolated subjects}

In a parallel experiment, ``isolated'' subjects received as social information an estimate generated from a random value of $\alpha$, uniformly distributed between $-3$ and 3.
The model for isolated subjects is very similar to the one presented above, and also sticks closely to the experimental structure: instead of receiving the average of $\tau$ previous estimates, agents receive a value of $\alpha$ randomly generated from a uniform distribution. 
There is no loop over $\alpha$ here, and the variables $\rho$ and $\tau$ do not exist in this separate experiment.
All the rest is the same as in the model of the main experiment.
10000 simulations were run with 200 agents in each.

\subsection*{Computation of the error bars}

The error bars indicate the variability of our results depending on the $N_{\rm Q} = 32$ questions presented to the subjects.
We call $x_0$ the actual measurement of a quantity appearing in the figures by considering all $N_{\rm Q}$ questions asked. Then, we generate the results of $N_{\rm exp} = 1000$ new effective experiments. For each effective experiment indexed by $j=1,...,N_{\rm exp}$, we randomly draw $N_{\rm Q}' = N_{\rm Q}$ questions among the $N_{\rm Q}$ questions  asked (so that some questions can appear several times, and others may not appear) and recompute the quantity of interest which now takes the value $x_j$. The upper error bar $b_+$ for $x_0$ is defined so that $C=68,3 \,$\% (by analogy with the usual standard deviation for a normal distribution) of the $x_j$ greater than $x_0$ are between $x_0$ and $x_0+b_+$. Similarly, the lower error bar $b_-$  is defined so that $C=68,3 \,$\% of the $x_j$ lower than $x_0$ are between $x_0-b_-$ and $x_0$. The introduction of these upper and lower confidence intervals is adapted in cases where the distribution of the $x_j$ is unknown and potentially not symmetric.

\end{document}


\maketitle

\newpage

\section{List of Questions}

We list below  the 32 questions used in the experiment and the corresponding correct answers $T$. In the original experiment, the questions were asked in French. 
Questions 1 to 29 were classified into 4 clear-cut categories, and questions 30 to 32 do not clearly belong to either of these categories. \\

\noindent \textbf{1. Visual perception (number or length of objects in an image):}

\begin{enumerate}	
		
	\item Marbles 1: How many marbles are in this jar? $T = 100$
	
	\includegraphics[width=0.35\textwidth]{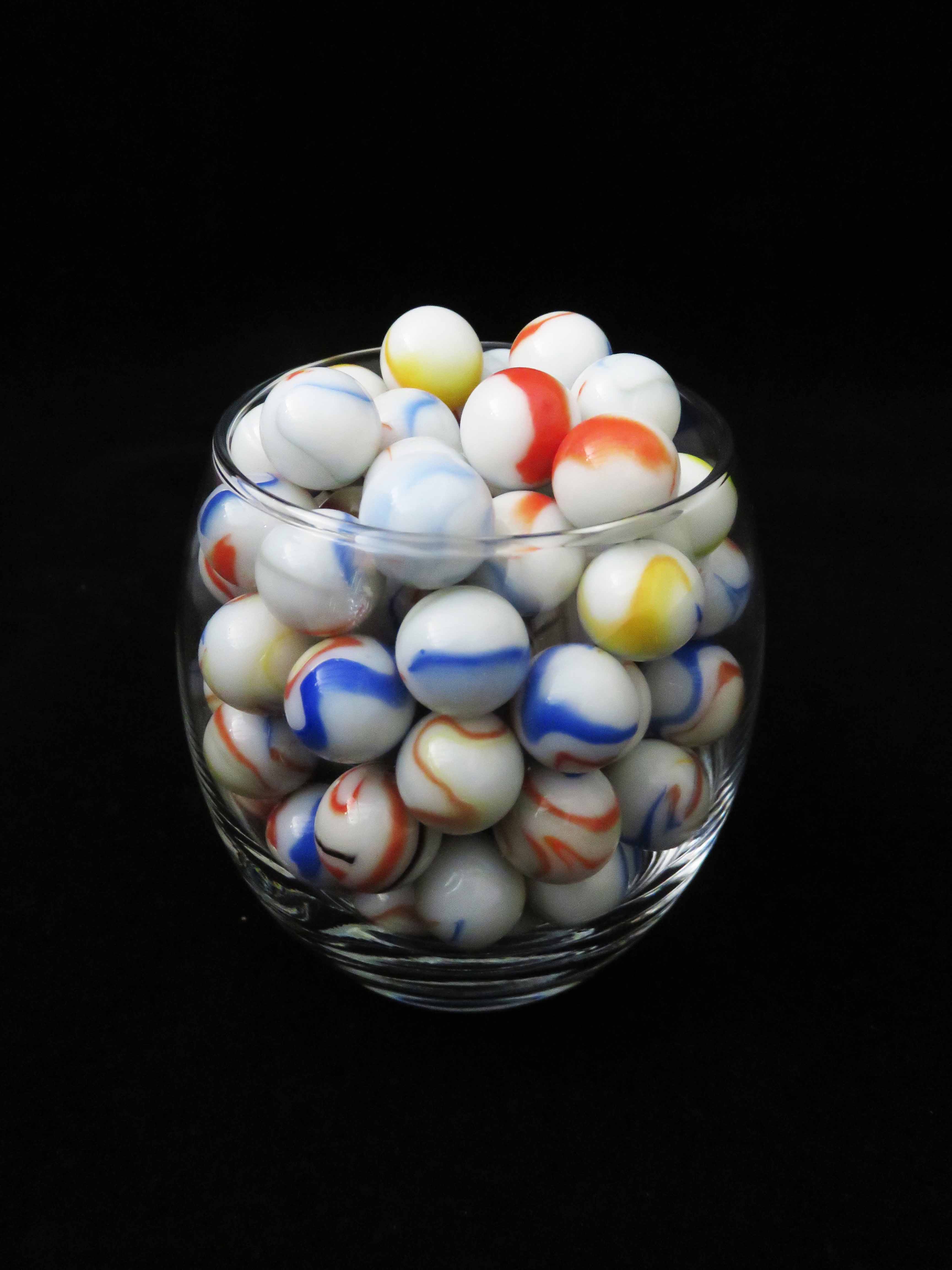}				
	
	\item Matches 1: How many matches can you see? $T = 240$
	
	\includegraphics[width=0.35\textwidth]{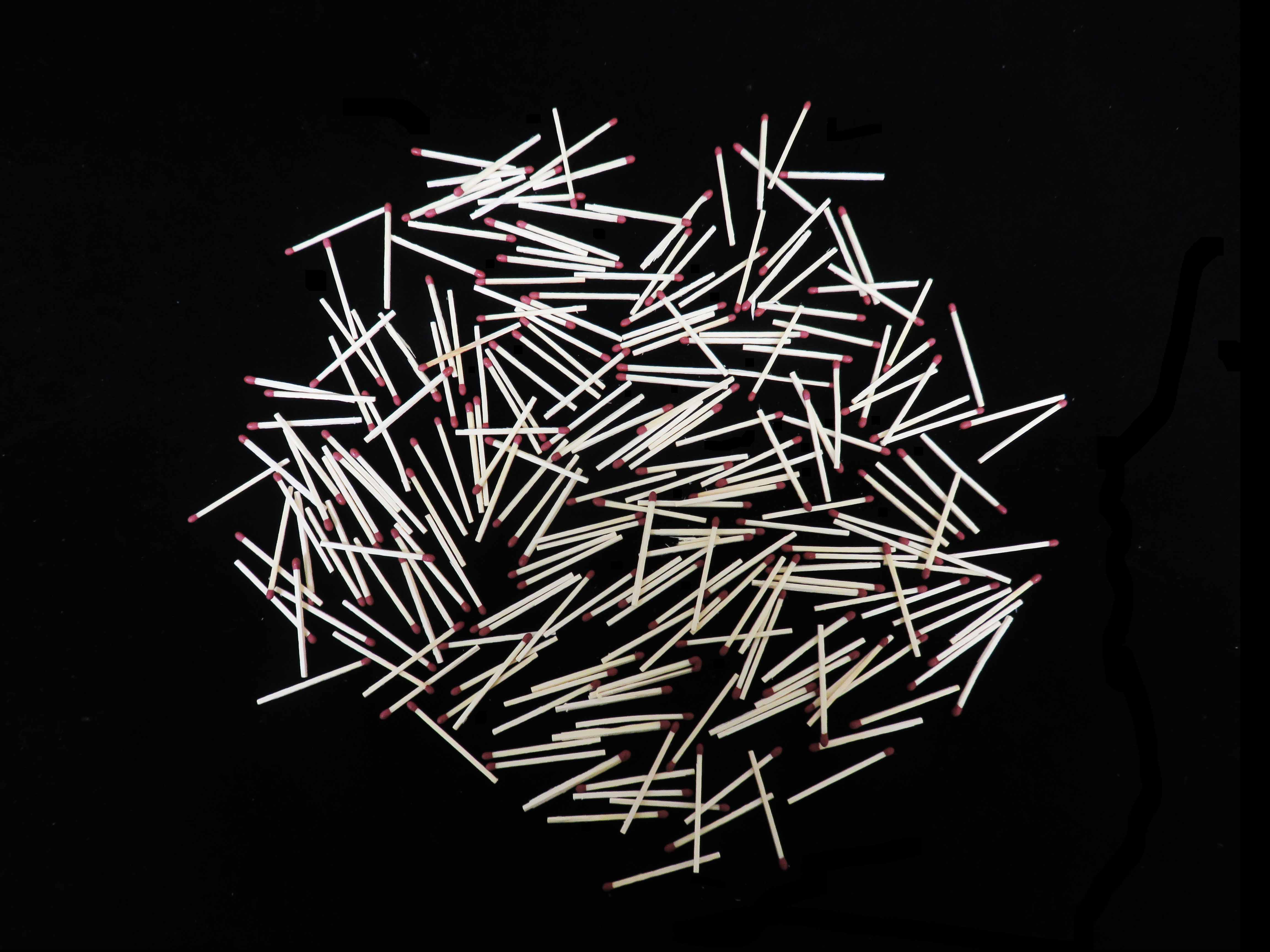}	
		
	\newpage	
	\item Matches 2: How many matches can you see? $T = 480$
	
	\includegraphics[width=0.31\textwidth]{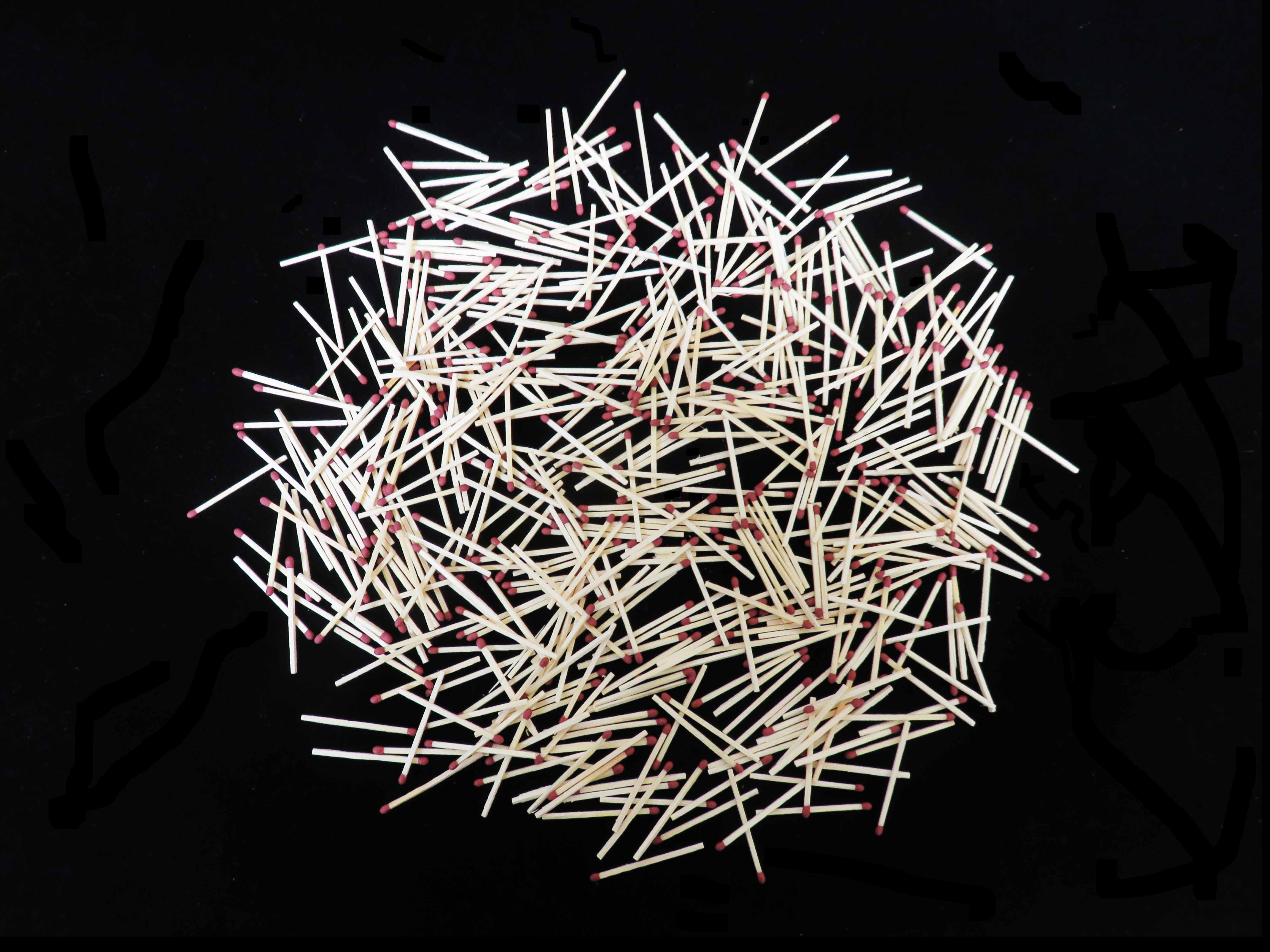}	
		
	\item Marbles 2: How many marbles are in this jar? $T = 450$
	
	\includegraphics[width=0.31\textwidth]{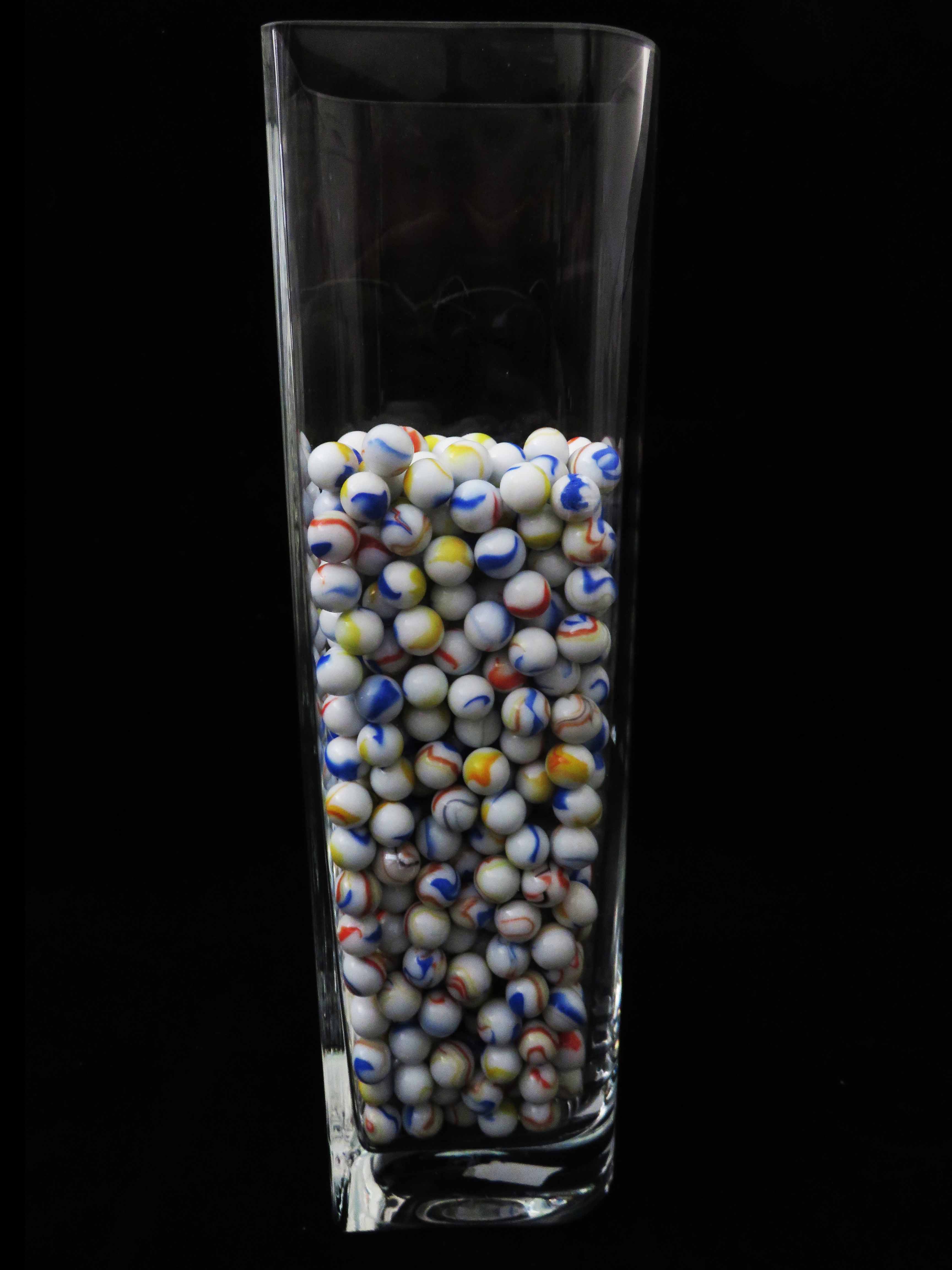}		
		
	\item Matches 3: How many matches can you see? $T = 720$
	
	\includegraphics[width=0.31\textwidth]{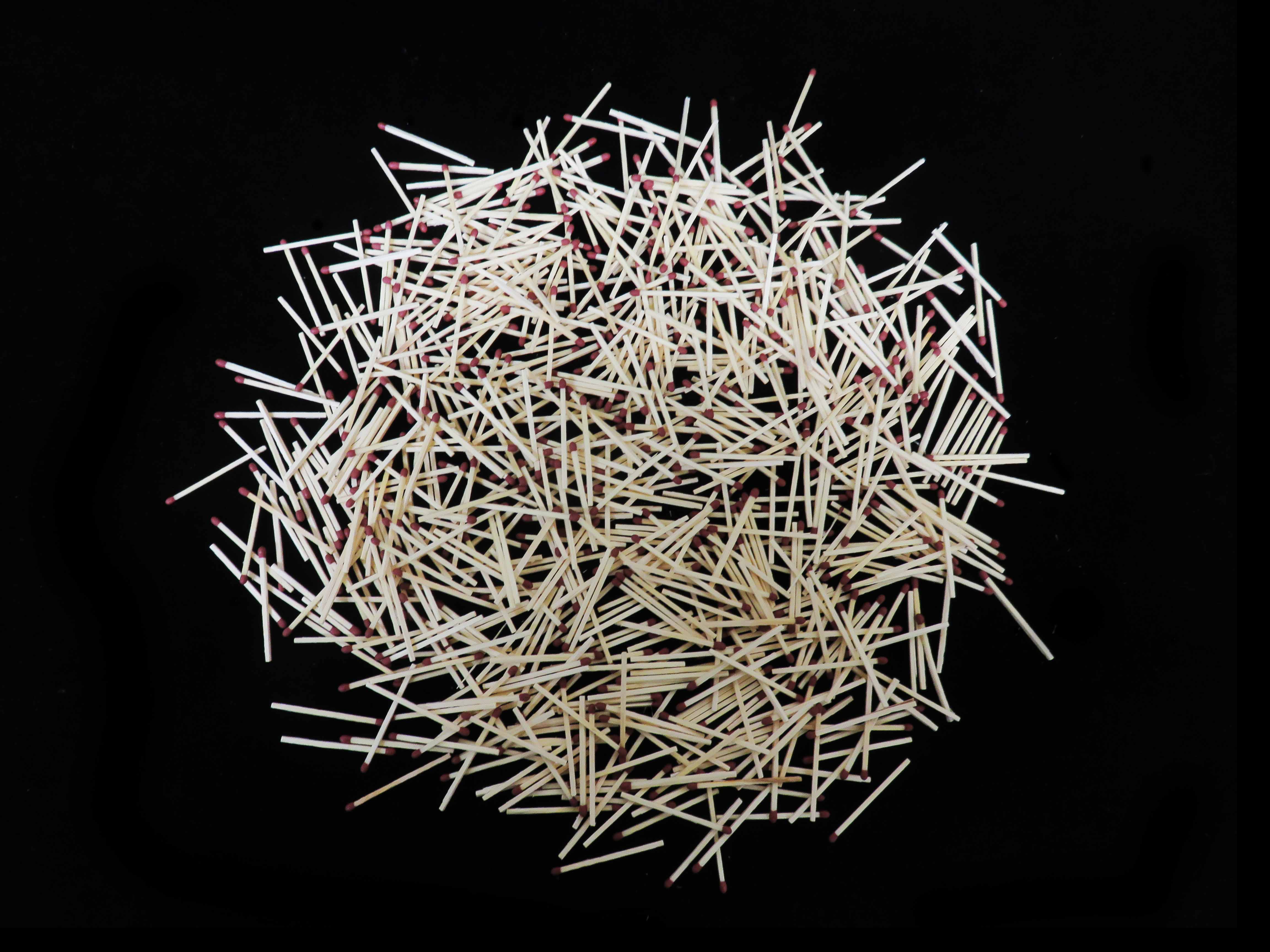}			
	
		\newpage	
	\item Rope 1: In your opinion, how long is this rope (in cm)? $T = 200$
	
	\includegraphics[width=0.31\textwidth]{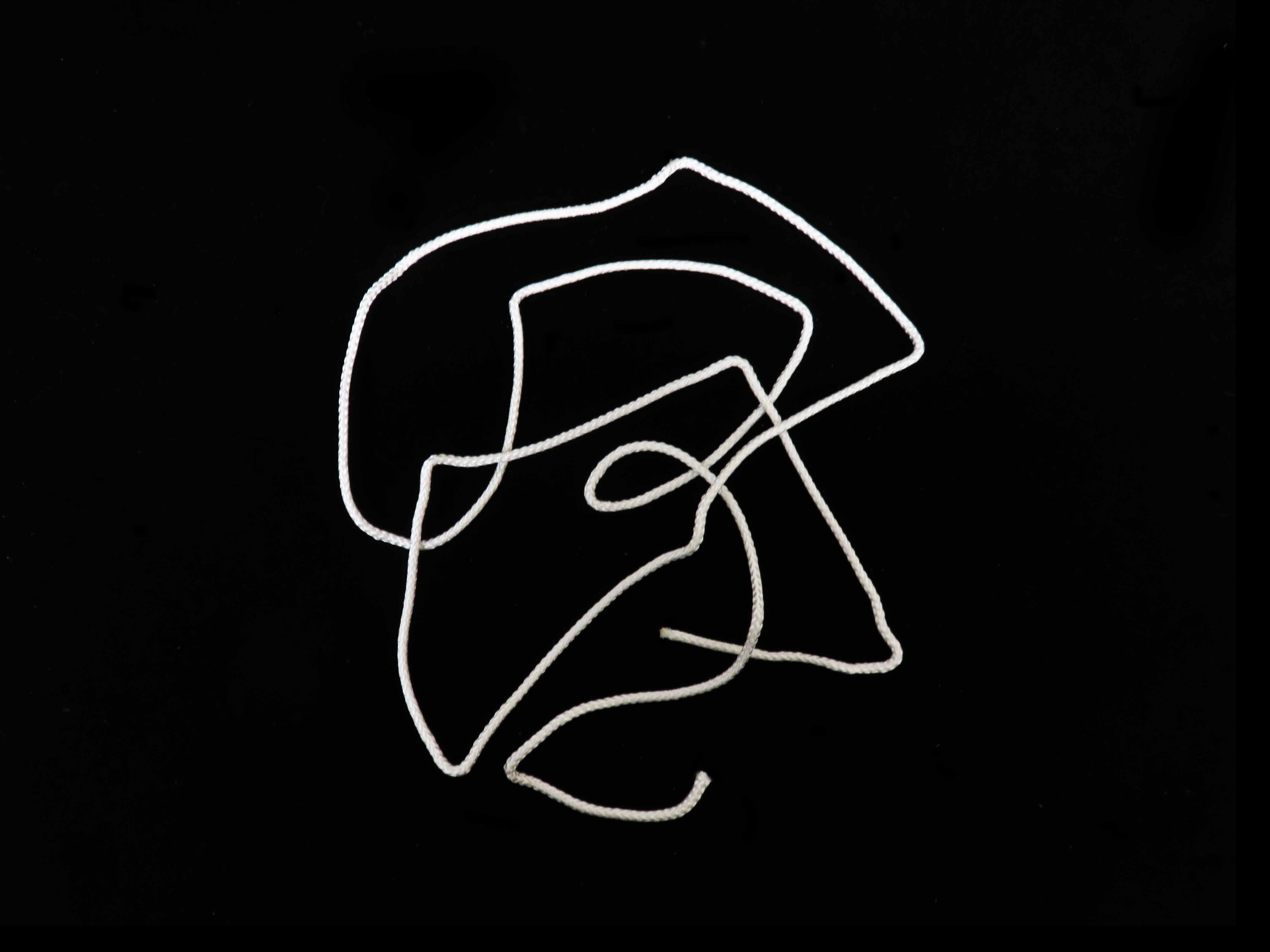}		 
			
	\item Rope 2: In your opinion, how long is this rope (in cm)? $T = 700$
	
	\includegraphics[width=0.31\textwidth]{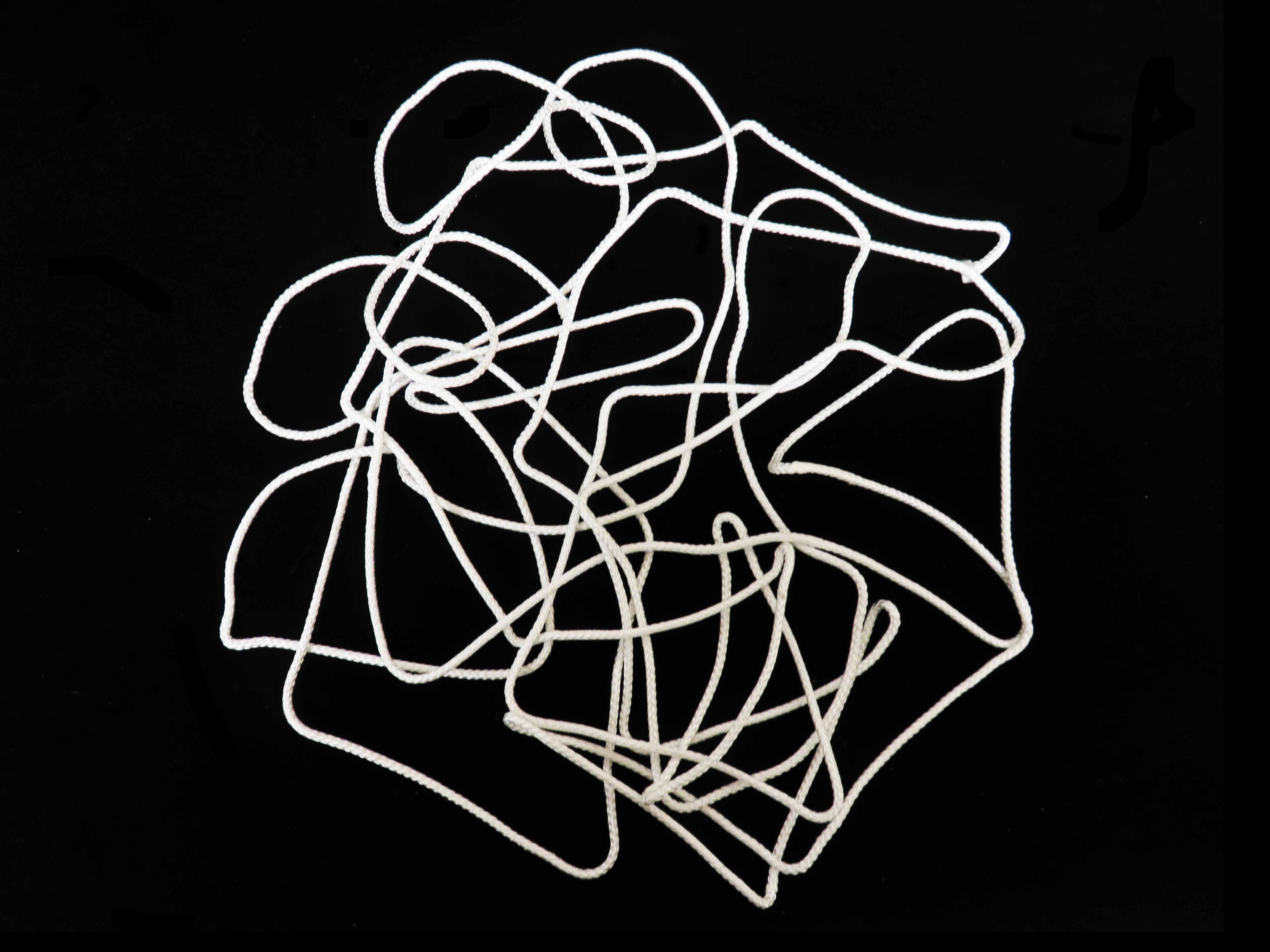}	
	
\end{enumerate}

\noindent \textbf{2. Population of large cities in the world:}

\begin{enumerate}

	\setcounter{enumi}{7}	
			
	\item What is the population of New York City and its agglomeration? $T = 21,000,000$
			
	\item What is the population of Madrid and its agglomeration? $T = 6,500,000$
	
	\item What is the population of Amsterdam and its agglomeration? $T = 1,600,000$
	
	\item What is the population of Tokyo and its agglomeration? $T = 38,000,000$
	
	\item What is the population of Melbourne and its agglomeration? $T = 4,500,000$
	
	\item What is the population of Seoul and its agglomeration? $T = 26,000,000$

	\item What is the population of Shanghai and its agglomeration? $T = 25,000,000$			
	
\end{enumerate}

\noindent \textbf{3. Daily life facts:}

\begin{enumerate}	
			
	\setcounter{enumi}{14}			
			
	\item How many kilometers does a professional cyclist typically bike a year? $T = 40,000$
	
	\item What is the mean annual gross salary of a professional league 1 soccer player in France (in euros)? $T = 600,000$
	
	\item How many cell phones are sold in France every year? $T = 22,000,000$	
	
	\item How many cars are stolen in France every year? $T = 110,000$		
			
	\item How many ebooks were sold in France in 2014? $T = 10,700,000$
			
	\item How many books does the American Library of Congress hold? $T = 23,000,000$
			
	\item How many people die from cancer in the world every year? $T = 15,000,000$
	
\end{enumerate}

\noindent \textbf{4. Extreme (astronomical or biological/geological) events:}

\begin{enumerate}	
	
	\setcounter{enumi}{21}	
	
	\item What is the radius of the Sun (in km)? $T = 696,000$
			
	\item What is the distance between the Earth and the Moon (in km)? $T = 385,000$
	
	\item What is the mean distance between planet Mercury and  the Sun (in km)? $T = 57,800,000$	
	
	\item What is the total mass of oceans on Earth (in thousand billion tons)? $T = 1,400,000$
	
	\item How many cells are there in the human body (in billion cells)? $T = 100,000$
	
	\item How many galaxies does the visible universe hold (in million galaxies)? $T = 100,000$
			
	\item How many stars does the Milky way hold (in million stars)? $T = 235,000$		
		
	\item How many billions kilometers is worth a light-year? $T = 9,461$
			
\end{enumerate}

\noindent \textbf{5. Other:	}	
				
\begin{enumerate}	
	
	\setcounter{enumi}{29}	
	
	\item What is the mass of the Cheops pyramid (in tons)? $T = 5,000,000$	
	
	\item What is the total length of the metal threads used in the Golden Gate Bridge's braided cables (in km)? $T = 129,000$		
		
	\item How much did Burj Khalifa Tower, in Dubai, cost to build (in thousand dollars)? $T = 1,500,000$						
		
\end{enumerate}


\section{Supplementary Figures}

\begin{figure} [!h]
\centering
	\includegraphics[width=1\textwidth]{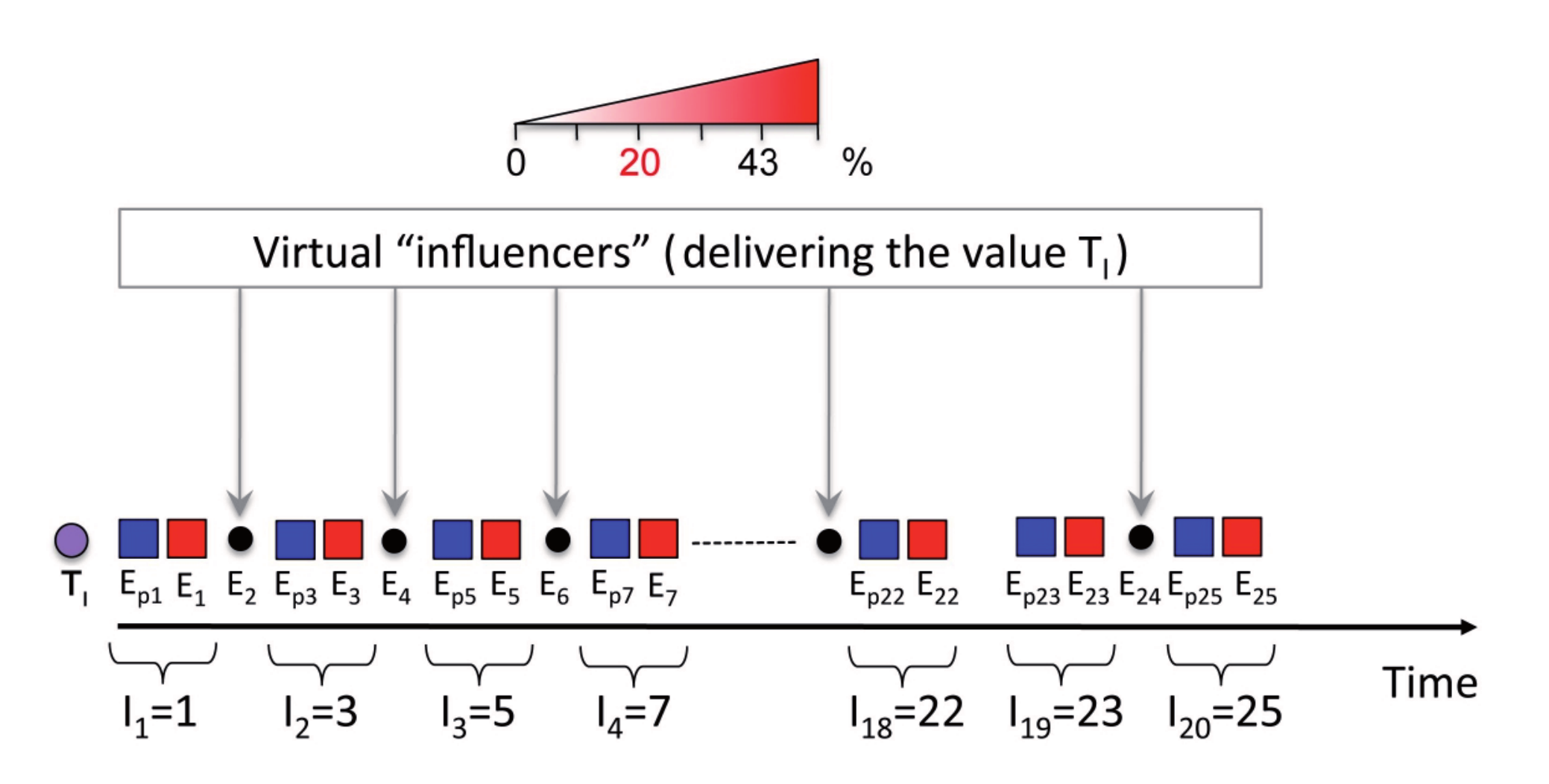}
		\caption{Sequence of estimates for a given question: one after the other, each individual $i_k$ ($k=1,..., 20)$ first gives his/her personal estimate $E_{\rm p}$ (in blue). Then, after receiving social information (the geometric mean of the $\tau$ previous estimates), he/she gives his/her second estimate (in red). $n = 0$, 5, or 15 \textit{virtual influencers} (black dots), corresponding to a fraction $\rho = \frac{n}{n+20} = 0$, 20, or $43\%$ respectively, are added at random locations in the sequence, without the subjects being aware of it. The value $T_{\rm I}$ provided by the \textit{virtual influencers} is chosen as initial condition (violet circle). 
		The figure shows an example with 5 virtual influencers, corresponding to a fraction $\rho = 20\%$ (highlighted in red on top of the panel). In this example, $\tau = 1$, such that subjects receive as social information the estimate of the subject or \textit{virtual influencer} that comes right before them in the sequence. For instance, subject $I_2$ receives $T_{\rm I}$ as social information from the first \textit{virtual influencer} ($E_2 = T_{\rm I}$), and subject $I_{19}$ receives $E_{22}$ as social information from subject $I_{18}$.}
		\label{figS1}
\end{figure}

\newpage

\begin{figure} [h!]
	\centering
	\includegraphics[width=1\textwidth]{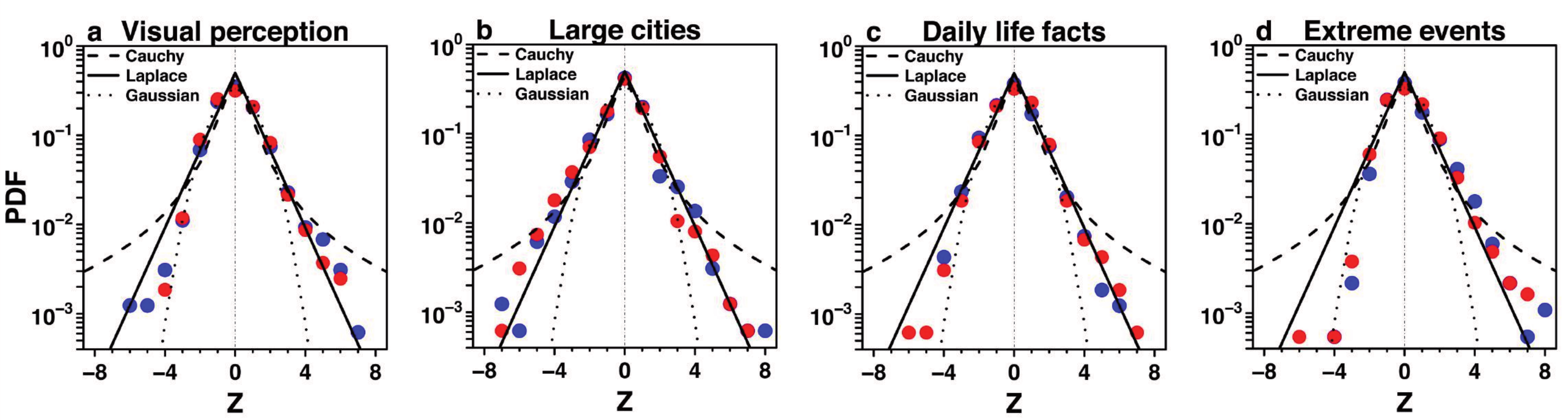}
	\caption{Distribution of fully normalized estimates $Z = \frac{X - m}{\sigma}$, before (blue) and after (red) social influence, for the four categories defined: (a) visual perception, (b) population of large cities in the world, (c) daily life events and (d) extreme (astronomical or biological/geological) events.
		}
		\label{figS2}
\end{figure}

\begin{figure} [h!]
	\centering
	\includegraphics[width=0.5\textwidth]{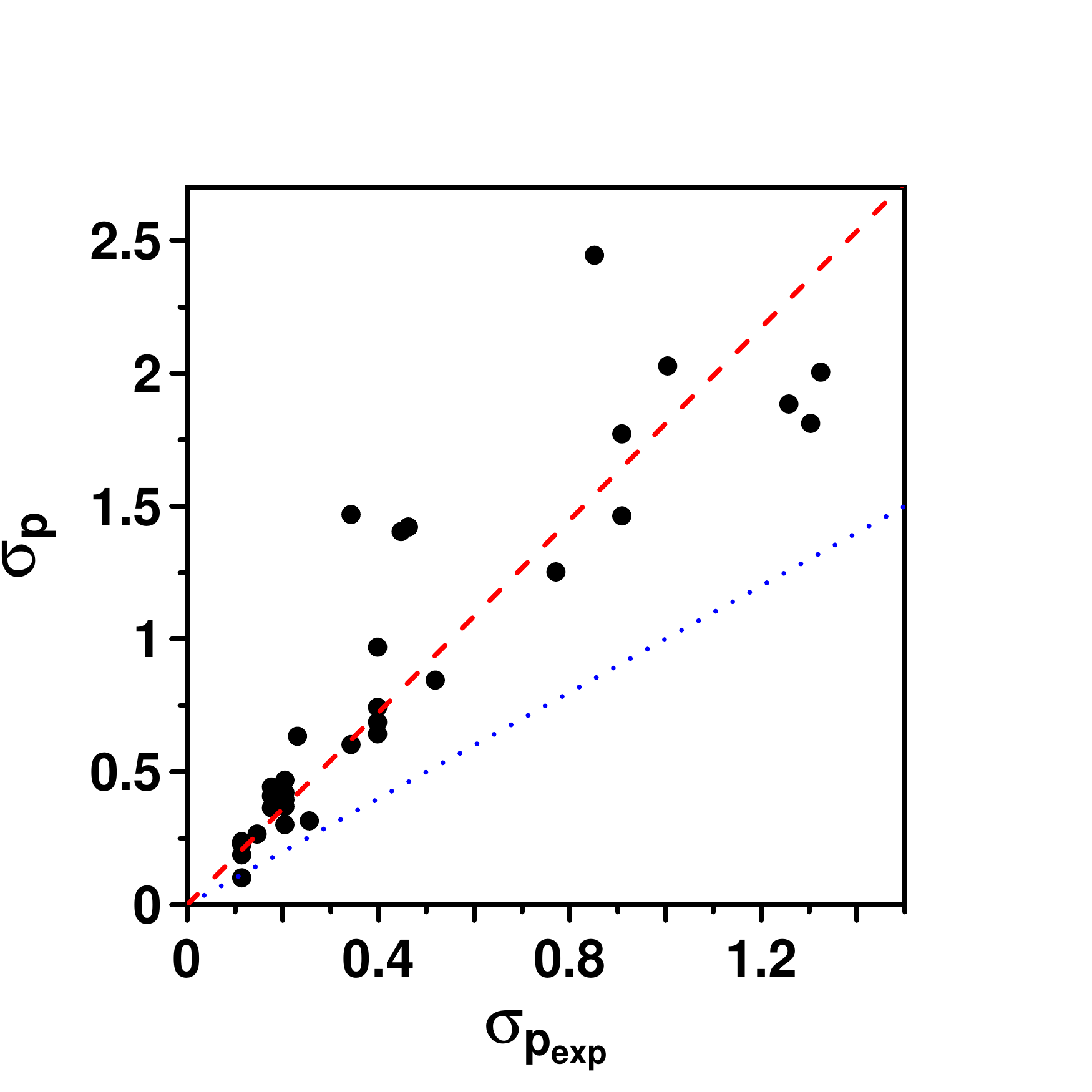}
	\caption{Dispersion $\sigma_{\rm p} = \langle |X_{\rm p} - m_{\rm p}| \rangle$ of estimates $X_{\rm p} = \log(\frac{E_{\rm p}}{T})$, for the 32 questions asked in the experiment, against the expected values of the dispersion ${\sigma_{\rm p}}_{\rm exp}$ used to define the information $T_{\rm I}$ provided by the influencers. The red dashed line is a linear regression of slope about $1.8$, while the blue dotted line has a slope of 1. The red line being above the blue line means that our expected values underestimated the actual dispersion of estimates.
		}
		\label{figS3}
\end{figure}

\newpage

\begin{figure} [h!]
	\centering
	\includegraphics[width=1\textwidth]{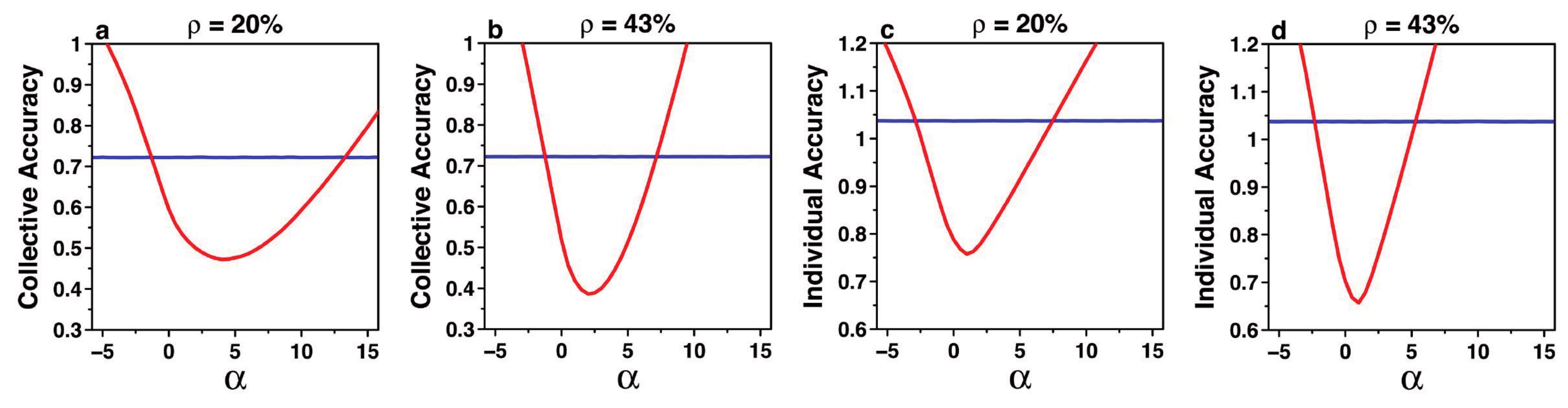}
	\caption{Model simulations of collective (a and b) and individual (c and d) accuracy, as a function of the quantifier of information quality $\alpha$, before (blue) and after (red) social influence, for $\rho = 20 \, \%$ (a and c) and $\rho = 43 \, \%$ (b and d) of \textit{virtual influencers} in the sequence of estimates. 
	Collective and individual accuracy can improve when \textit{virtual influencers} provide information that overestimate the truth by far, especially when $\rho = 20 \,\%$, but decrease sharply when \textit{virtual influencers} provide information that underestimate the truth: reinforcing the group bias has a strong negative impact on its accuracy.}
	\label{figS4}
\end{figure}

\begin{figure} [h!]
	\centering
	\includegraphics[width=1\textwidth]{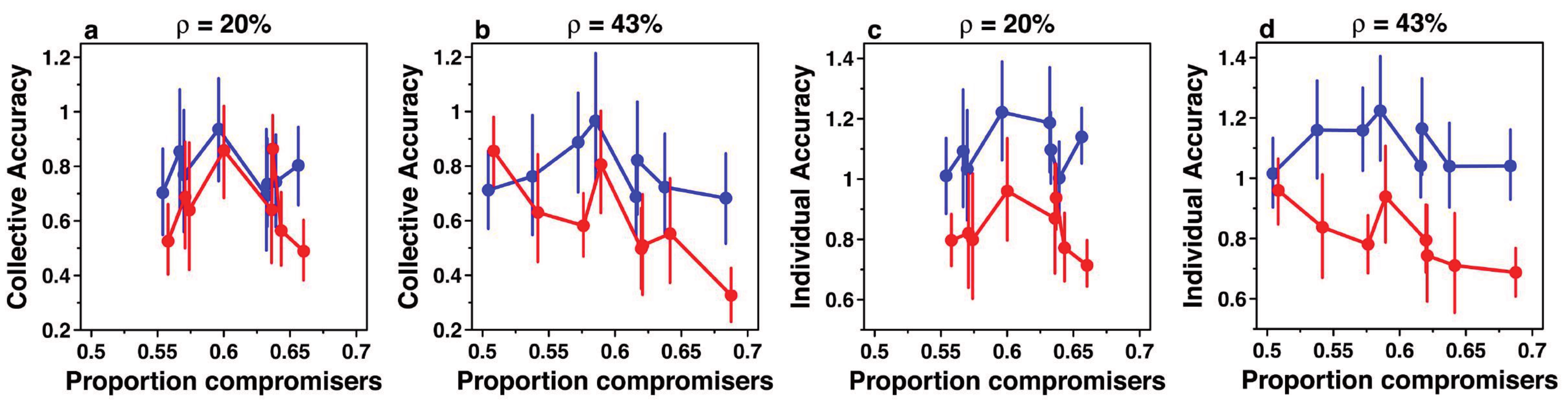}
	\caption{Collective (a and b) and individual (c and d) accuracy, as a function of the proportion of compromisers, before (blue) and after (red) social influence, for $\rho = 20 \, \%$ (a and c) and $\rho = 43 \, \%$ (b and d) of influencers in the sequence of estimates. Each dot corresponds to a value of $\alpha$.
	When $\rho = 43\,\%$, the individual and collective accuracies significantly improve (i.e., gets closer to 0) after social influence (red) with an increasing proportion of compromisers, but not when $\rho = 20\,\%$.}
	\label{figS5}
\end{figure}

\newpage

\section{Supplementary Table}

\begin{table} [h]
\rowcolors{2}{gray!25}{white}
\centering
\scalebox{0.88}{%
\begin{tabular}{| c | c | c | c | c | c | c | c | c | c | c | c |}
\hline
\multicolumn{3}{|c|}{} & \multicolumn{4}{c}{$\rho = 20 \, \%$} & \multicolumn{4}{|c|}{$\rho = 43 \, \%$} \\ \hline
Question & $\tau$ & ${\sigma_{\rm p}}_{\rm exp}$ & Subj. 2 & Subj. 3 & Subj. 4 & Subj. 5 & Subj. 6 & Subj. 7 & Subj. 8 & Subj. 9 \\ \hline

1 & 1 & 0.114 & 0.5 & -0.5 & 1   & -1   & 0.5 & -0.5 & 1   & -1    \\ \hline
2 & 1 & 0.114 & 2   & -2   & 3   & 1.5  & 2   & -2   & 3   & 1.5   \\ \hline
3 & 3 & 0.114 & 3   & 1.5  & 2   & -2   & 3   & 1.5  & 2   & -2    \\ \hline
4 & 3 & 0.114 & 1   & -1   & 0.5 & -0.5 & 1   & -1   & 0.5 & -0.5  \\ \hline
5 & 1 & 0.146 & 2   & -2   & 3   & 1.5  & 2   & -2   & 3   & 1.5   \\ \hline
6 & 1 & 0.204 & 0.5 & -0.5 & 2   & -2   & 0.5 & -0.5 & 2   & -2    \\ \hline
7 & 3 & 0.255 & 1   & -1   & 3   & 1.5  & 1   & -1   & 3   & 1.5   \\ \hline
8  & 3 & 0.176 & 3   & 1.5  & 2   & -2   & 3   & 1.5  & 2   & -2    \\ \hline
9  & 1 & 0.204 & 0.5 & -0.5 & 2   & -2   & 0.5 & -0.5 & 2   & -2    \\ \hline
10 & 1 & 0.204 & 0.5 & -0.5 & 1   & -1   & 0.5 & -0.5 & 1   & -1    \\ \hline
11  & 1 & 0.176 & 0.5 & -0.5 & 1   & -1   & 0.5 & -0.5 & 1   & -1    \\ \hline
12  & 3 & 0.204 & 1   & -1   & 3   & 1.5  & 1   & -1   & 3   & 1.5   \\ \hline
13  & 1 & 0.176 & 2   & -2   & 3   & 1.5  & 2   & -2   & 3   & 1.5   \\ \hline
14  & 3 & 0.204 & 1   & -1   & 0.5 & -0.5 & 1   & -1   & 0.5 & -0.5  \\ \hline
15 & 1 & 0.342 & 2   & -2   & 0.5 & -0.5 & 2   & -2   & 0.5 & -0.5  \\ \hline
16 & 3 & 0.230 & 1   & -1   & 0.5 & -0.5 & 1   & -1   & 0.5 & -0.5  \\ \hline
17 & 3 & 0.398 & 1   & -1   & 2   & -2   & 1   & -1   & 2   & -2    \\ \hline
18 & 3 & 0.398 & 3   & 1.5  & 1   & -1   & 3   & 1.5  & 1   & -1    \\ \hline
19  & 1 & 0.398 & 2   & -2   & 1   & -1   & 2   & -2   & 1   & -1    \\ \hline
20  & 3 & 0.519 & 3   & 1.5  & 0.5 & -0.5 & 3   & 1.5  & 0.5 & -0.5  \\ \hline
21  & 1 & 0.398 & 0.5 & -0.5 & 3   & 1.5  & 0.5 & -0.5 & 3   & 1.5   \\ \hline
22 & 1 & 0.447 & 2   & -2   & 1   & -1   & 2   & -2   & 1   & -1    \\ \hline
23 & 3 & 0.462 & 3   & 1.5  & 0.5 & -0.5 & 3   & 1.5  & 0.5 & -0.5  \\ \hline
24 & 1 & 0.908 & 0.5 & -0.5 & 2   & -2   & 0.5 & -0.5 & 2   & -2    \\ \hline
25 & 3 & 1.324 & 3   & 1.5  & 2   & -2   & 3   & 1.5  & 2   & -2    \\ \hline
26 & 3 & 1.258 & 3   & 1.5  & 1   & -1   & 3   & 1.5  & 1   & -1    \\ \hline
27 & 1 & 1.324 & 2   & -2   & 0.5 & -0.5 & 2   & -2   & 0.5 & -0.5  \\ \hline
28 & 1 & 1.004 & 0.5 & -0.5 & 3   & 1.5  & 0.5 & -0.5 & 3   & 1.5   \\ \hline
29 & 3 & 0.851 & 1   & -1   & 2   & -2   & 1   & -1   & 2   & -2    \\ \hline
30 & 1 & 0.771 & 2   & -2   & 1   & -1   & 2   & -2   & 1   & -1    \\ \hline
31 & 3 & 0.908 & 1   & -1   & 3   & 1.5  & 1   & -1   & 3   & 1.5   \\ \hline
32 & 3 & 0.324 & 3   & 1.5  & 0.5 & -0.5 & 3   & 1.5  & 0.5 & -0.5  \\ \hline
\end{tabular}
}
\caption{Parameter values in a given session: each subject in a session is associated a specific value of $\rho$ for all questions, but the values of $\tau$ and $\alpha$ vary across questions. The ``Subjects'' columns (Subj. $i$, $i = 2$ ... 9) give the values of $\alpha$ for each question.} \label{main_table}
\end{table}